\documentclass{aa}

\usepackage{amsmath, amssymb, amsfonts, amscd}
\usepackage{mathtools} 
\usepackage{array} 
\usepackage[varg]{txfonts} 
\usepackage{bm}

\usepackage{graphicx} 

\usepackage{natbib}
\bibpunct{(}{)}{;}{a}{}{,} 

\begin{document}

\title{Parametrisation of the wave - zonal flow interactions\\ taking into account the full Coriolis acceleration}
\subtitle{The necessity of going beyond the traditional approximation\\ in the sub-inertial regime and in weakly stratified regions
}

\author{S. Mathis\inst{1}}

 \institute{Universit\'e Paris-Saclay, Universit\'e Paris Cit\'e, CEA, CNRS, AIM, F-91191 Gif-sur-Yvette, France\\
 \email{stephane.mathis@cea.fr}}

\date{Received ... / accepted ...}

\abstract 
{From the Earth's atmosphere and oceans to stellar radiation zones, inertia-gravity waves, which are called gravito-inertial waves (hereafter GIWs) in Astrophysics, are transporting momentum and mixing matter when they are damped through heat and viscous diffusions and when they break. Their short-time scale dynamics is governed by the combined action of the buoyancy force and of the Coriolis acceleration. Because of the transport they trigger, they modify the long-term evolution of the large-scale planetary atmospheric (oceanic) circulation and of the structure and rotation of stars. In many state-of-the-art models, the so-called Traditional Approximation of Rotation (hereafter denoted TAR), where the local projection of the rotation vector along the horizontal direction is neglected, is assumed.}
{The TAR can be adopted in the cases of very thin fluid layers or when the projection of the Coriolis acceleration along the direction of the entropy and chemical stratifications can be neglected when compared to the buoyancy force. It is often assumed when evaluating the fluxes of momentum, heat and matter transported by GIWs. We aim to identify the applicability regime of this approximation and to propose a non-traditional parametrisation of wave - zonal flow interactions in planetary atmospheres (oceans) and stellar interiors, in which the full Coriolis acceleration is taken into account.}
{We build a prototype local non-traditional Cartesian model in which we take into account the full Coriolis acceleration, buoyancy, and heat and viscous diffusions. We study the two channels through which GIWs are exchanging momentum with mean flows while transporting heat and mixing matter: their linear damping and their nonlinear breaking. We don't assume any hiearchy between stratification and rotation to allow us to explore the possible large parameter space in geophysical and astrophysical flows, in particular the cases of weak stable stratification and rapid rotation.}
{On the one hand, the radiative and viscous dampings of GIWs are increasing with a decreasing ratio of the wave frequency ($\omega$) with the inertial frequency ($2\Omega$, $\Omega$ being the angular rotation frequency). In the sub-inertial regime ($\omega<2\Omega$), the TAR is stongly underestimating GIWs damping, especially in the case where the buoyancy and the Coriolis acceleration have close strengths. In this regime a non-traditional modelling must be adopted. It predicts the correct altitude where momentum is deposited, which is closer to the excitation region of waves than the one predicted using the TAR. On the other hand, non-traditional modellings of GIWs convective and shear-induced breakings are proposed and we demonstrate that the TAR is overestimating the flux of momentum deposited by GIWs through these channels. Taking into account the full Coriolis acceleration leads to a stronger inhibition of the efficiency of the convective and shear-induced overturnings and to a weaker transport than those predicted when assuming the TAR. Finally, a fully non-traditional parametrisation of GIWs - mean zonal flows interaction is derived, which can be implemented in numerical models of the long-term evolution of the atmospheric (oceanic) general circulation and of the structure of rotating stars.} 
{}


\titlerunning{}
\authorrunning{Mathis}

\maketitle


\section{Introduction and context}
From the Earth's atmosphere and oceans to the radiative core of solar-type stars and the radiative envelope of early-type stars, the combined action of the buoyancy force, of the Coriolis acceleration, and of heat and viscous diffusions plays a key role in wave-driven momentum transport and chemical mixing  \citep[e.g.][]{Lindzen1981,VF2005,LottGuez2013,MacKinnonetal2017,Sutherlandetal2019,Ribsteinetal2022,Achatzetal2024,CT2005,TC2005,Rogersetal2013,Rogers2015,RE2017,Pinconetal2017,Vargheseetal2024}. In this framework, inertia-gravity waves \citep[called gravito-inertial waves in astrophysics, hereafter GIWs;][]{DintransRieutord2000}, which have buoyancy and the Coriolis acceleration as restoring forces, deposit their angular momentum at the place where they are damped \citep[e.g.][]{Zahnetal1997,VF2005}, where they encounter critical layers \citep[e.g.][]{BookerBretherton1967,Alvanetal2013,Lottetal2015} and where they break \citep[e.g.][]{Sutherland2001,StaquetAR2002,Rogersetal2013,Mathis2025}. If many works have focused on simple gravity waves \citep[e.g.][]{Press1981,Schatzman1993,Zahnetal1997}, the Coriolis acceleration must be taken into account in many geophysical and astrophysical configurations such as in the case of weakly stratified oceanic layers \citep[e.g.][]{GS2005,Gerkemaetal2008,ShriraTownsend2010} and in the case of rapidly-rotating young low-mass stars and early-type stars \citep[e.g.][]{Pantillonetal2007,Mathisetal2008}. A key theoretical challenge must then be addressed in the case when the buoyancy frequency ($N$, also called the Brunt-V\"ais\"al\"a frequency) and the inertia frequency ($2\Omega$, $\Omega$ being the rotation) are of the same order of magnitude, i.e. $N\sim{\mathcal C}\left(2\Omega\right)$ with $1<{\mathcal C}<10$. In this regime, the propagation of GIWs must be modelled in a bi-dimensional non-separable mathematical framework \citep[e.g.][]{DintransRieutord2000,GS2005}. This is a challenging situation both theoretically and numerically.\\

In this context, most of the work studying transport and mixing by GIWs in a geophysical or astrophysical context have adopted the so-called Traditional Approximation of Rotation (herafter denoted TAR) where the local projection of the rotation vector along the latitudinal direction is neglected \citep[e.g.][]{Eckart1961,Bildstenetal1996,LeeSaio1997,Mathisetal2008,Mathis2009,Leeetal2014}. It involves neglecting the component of the Coriolis acceleration along the direction of the entropy and chemical stratifications that allows us to uncouple the horizontal and the vertical dynamics and a separation of variables in the mathematical modelling of GIWs. This simplified mathematical formalism can be applied when $2\Omega\!<\!\!<\!N$ but leads to erroneous predictions when $2\Omega\sim N$ for GIWs propagation \citep[e.g.][]{Friedlander1987,GS2005,Fruman2009,Pratetal2016,Pratetal2018}, hydrodynamical instabilities \citep[e.g.][]{Zeitlin2018,Parketal2021,TB2022,Ray2023,ParkMathis2025} and excitation of GIWs \citep[e.g.][]{MNT2014,Augustsonetal2020}.\\

In a situation where geophysical and astrophysical fluid dynamicists have to evaluate and parametrise the impact of GIWs on the large-scale atmospheric and oceanic circulation and its long term evolution and on the secular structural, chemical and dynamical evolution of stars, it is thus mandatory to evaluate when a non-traditional (herafter NT in the equations) modelling must be adopted and when the TAR can be used. This is the objective of this work, where we design a local Cartesian prototype model presented in \S2 to allow us to explore the linear damping of GIWs (in \S3) and their breaking (in \S4) for any $N/\left(2\Omega\right)$ ratio. In \S5, we provide the corresponding non-traditional parametrisation for GIWs-mean flows interactions and we discuss perspectives and conclusions of this work in \S6.

\section{Gravito-inertial waves and transport}
\label{giw}

In this work, our goal is to provide a global understanding of the effect of the complete Coriolis acceleration on the wave-mean flows interactions in stellar, planetary and geophysical flows. We will focus on the two channels that are driving these interactions: the linear damping of waves and their nonlinear breaking.

\subsection{A non-traditional prototype model}

\begin{figure}[t!]
\centering
\includegraphics[width=0.425\textwidth]{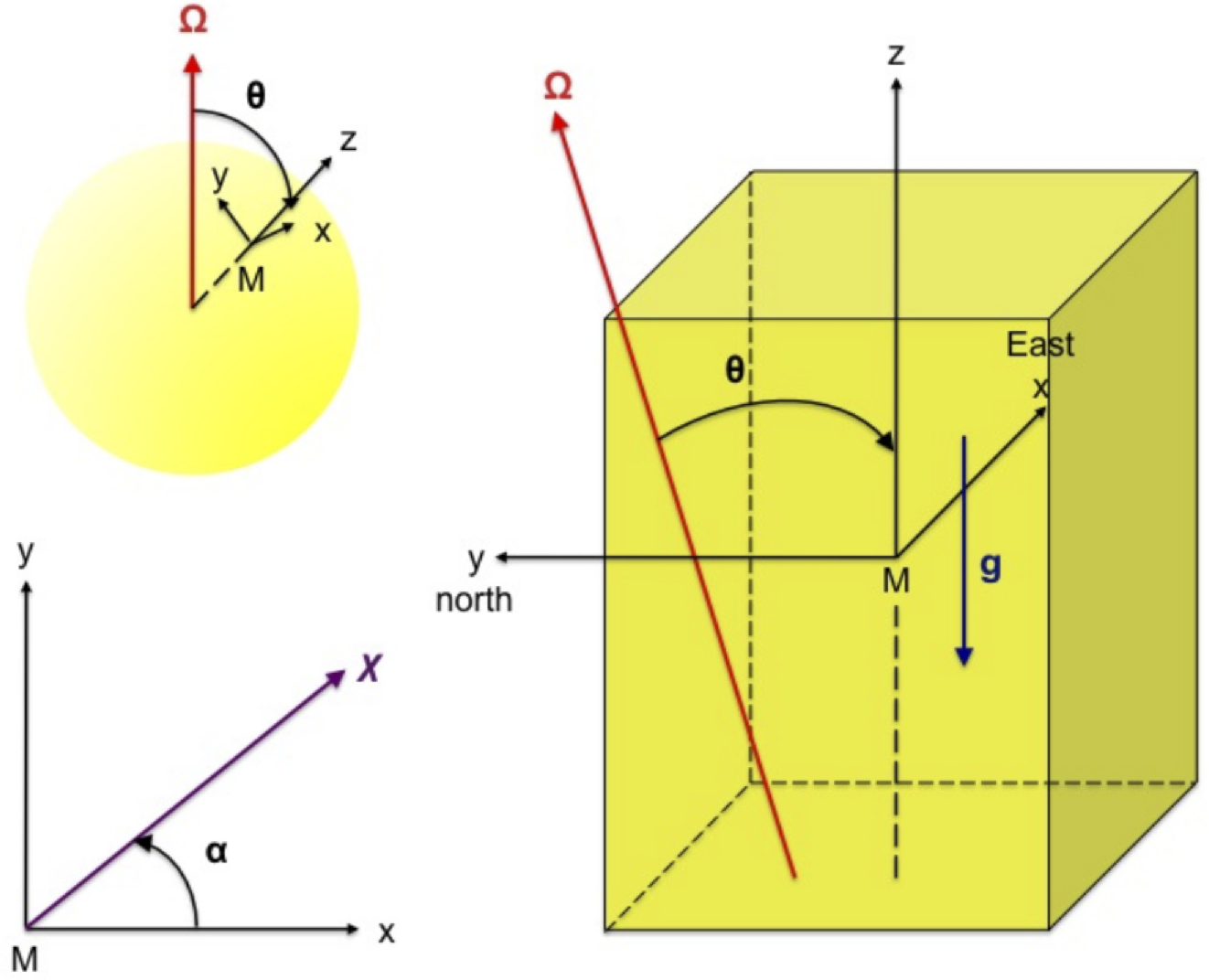}        
\caption{The local studied Cartesian box in the "$f$-plane" reference frame.}
\label{Fig1M}
\end{figure}

To reach this objective, we choose to adopt a local approach that allows us to unravel these physical mechanisms step by step. Therefore, we choose to consider a Cartesian box, centered on a point M with spherical coordinates $\left(r,\Theta,\varphi\right)$ inside a stably stratified rotating stellar or planetary region, where $r$ is the radius, $\Theta$ the angle between the local effective gravity ${\vec g}_{\rm eff}$\footnote{The effective gravity is the sum of the self-gravity ${\vec g}$ and of the centrifugal acceleration $1/2\,\Omega^2{\vec\nabla}s^2$, where $s$ is the distance from the rotation axis.} and the rotation vector $\vec\Omega$ (see Fig.~\ref{Fig1M}), and $\varphi$ is the azimuthal angle. M$x$, M$y$ and M$z$ are the axes along the local longitudinal (azimuthal), latitudinal and vertical (along ${\vec g}_{\rm eff}$) directions, respectively with the corresponding Cartesian coordinates $\left\{x,y,z\right\}$ and unit-vector basis $\left\{{\widehat {\bf e}}_{x},{\widehat {\bf e}}_{y},{\widehat {\bf e}}_{z}\right\}$. We introduce a reduced horizontal coordinate, $\chi$, that makes an angle $\alpha$ with respect to the $x-$axis: $\chi =x\cos\alpha + y\sin\alpha$. This so-called ``$f$-plane'' \citep{Pedlosky1982} is co-rotating with the rotation $\vec\Omega=\Omega\,{\widehat{\bf e}}_{\Omega}$, where $\Omega$ is the global mean planetary/stellar rotation with ${\widehat {\bf e}}_{\Omega}=\sin\Theta\,{\widehat {\bf e}}_{y}+\cos\Theta\,{\widehat {\bf e}}_z$. 

To model mean longitudinally averaged zonal flow (differential rotation), we follow \cite{Mathis2009} and we introduce in an inertial frame:
\begin{equation}
{\vec V}_{\rm 0}=\left({V}_{\Omega}+{U}\left(y,z\right)\right){\bf e}_{x},
\label{eq:ZFdef}
\end{equation}
where $V_{\Omega}=r\sin\theta\,\Omega$ is the zonal flow associated to the global mean rotation $\Omega$ at the point $M$ with spherical coordinates $\left\{r,\theta\right\}$ and ${U}$ is the local longitudinally averaged zonal flow, which depends on both the local latitudinal ($y$) and vertical ($z$) coordinates in the general case. In this work, we consider as a first step a local horizontally averaged zonal flow ${\overline U}\left(z\right)$, which depends only on $z$ and is the local equivalent of the so-called "shellular" differential rotation introduced by \cite{Zahn1992} in the global spherical case.\\

To treat the Coriolis acceleration in the equations of GIWs' dynamics written in this reference frame, we introduce the two components of $2{\vec\Omega}$\begin{equation}
f=2\Omega\cos\Theta\hbox{ }\hbox{ }\hbox{and}\hbox{ }\hbox{ }{\widetilde f}=2\Omega\sin\Theta,
\end{equation} 
along the vertical and latitudinal directions, respectively. In the most general case, both components must be taken into account to obtain a correct treatment of GIWs' dynamics \citep{GS2005,Gerkemaetal2008,MNT2014}. Then, assuming a constant $f$ leads to the so-called ``non-traditional $f$-plane'', contrary to the usual traditional approximation in which the horizontal component $\widetilde f$ is neglected \citep[][]{Eckart1961}. The validity of this approximation will be discussed throughout this work.

This local approach is valid only for mechanisms that can be modelled locally in a box such that
\begin{equation}
\left\{L_x, L_y, L_z\right\}\!<\!\!<\!R,
\label{cond:cartesianBoxSize1}
\end{equation}
where $L_x$, $L_y$, $L_z$ are the lengths of the box in the $x$, $y$, and $z$ directions, respectively, and $R$ is the radius of the body. In this framework, the linear damping of GIWs and their potential nonlinear breaking are maximum near the so-called critical layer where their local Doppler-shifted frequency ${\widehat \omega}=\omega+k_{x}{\overline U}$ ($\omega$ being the frequency of the studied GIW, $k_x$ its azimutal wavenumber with its phase $\left(\omega\,t+k_{x}x\right)$, and ${\overline U}$ the local horizontally-averaged (vertically-sheared) zonal flow introduced above) vanishes and where angular momentum is deposited or extracted \citep[][]{Alvanetal2013}. In our prototype non-traditional model, we consider that such a critical layer is within our box and that the wave-mean zonal flow interactions occur on spatial scales such that Eq. (\ref{cond:cartesianBoxSize1}) is verified. In our model, we also assume the Cowling approximation \citep{Cowling1941}, in which the fluctuations of the self-gravity induced by GIWs are neglected; this approximation is verified in the case of GIWs that are propagating towards their critical layer, which are rapidly-oscillating along the vertical direction. In the case where their local vertical wavelength is smaller than the local density and pressure height-scales ($H_{\rm \rho}\left(z\right)=\vert{\rm d}z/{\rm d}\ln {\overline \rho}\left(z\right)\vert$ and $H_{\rm p}\left(z\right)=\vert{\rm d}z/{\rm d}\ln {\overline P}\left(z\right)\vert$, ${\overline \rho}$ and ${\overline P}$ being the hydrostatic density and pressure, respectively), we can assume locally the Boussinesq approximation where the density fluctuations are only taken into account in the buoyancy force. Such a theoretical framework is coherent with the one that has been adopted in the case of stars for the study of the transport of angular momentum by low-frequency progressive internal gravity waves, which in most of the literature is based on the seminal work by \cite{Zahnetal1997}. In this work, the so-called quasi-adiabatic approximation is adopted \citep[see also][in the case of the Earth's atmosphere]{VF2005}: because of the small values of the viscosity and the heat diffusion in stars and in other celestial bodies, the damping of the waves they induce along their propagation is treated as a first-order local perturbation of their adiabatic propagation. Their adiabatic propagation is generally computed assuming the anelastic approximation, which allows us to take into account high density contrasts in stars and planetary atmospheres while filtering higher-frequencies acoustic waves \citep[we refer the reader to][for the case of planetary atmospheres]{Lindzen1981,ADML2018}, but their linear damping is computed locally using the Boussinesq approximation as we propose here. Such a simplification can be done because the terms of the heat and viscous diffusions, which scale with the Laplacian applied to the components of the velocity of the waves and their entropy and temperature fluctuations, and therefore as the square of the total wave vector, are dominating the terms relative to the variations of the background hydrostatic quantities.\\

In the forthcoming section \S2.2, we therefore derive the dissipative Poincar\'e equation, which describes the non-adiabatic propagation of GIWs in a local Boussinesq model. Next, in \S2.3, we recall the main key results on their adiabatic propagation, which will be used to study their linear damping in \S3 and their convective and shear-induced breaking in \S4.

\subsection{The dissipative Poincar\'e equation}

To derive the Poincar\'e equation that governs GIWs' dynamics, we write the linearised equations of motion of the studied stably stratified rotating stellar or planetary fluid in our local Cartesian set-up assuming the Boussinesq and the Cowling approximations as discussed in the previous section. We introduce the GIWs' velocity field $\vec u=\left(u,v,w\right)={\vec u}_H+{\vec u}_V$, where $u$, $v$ and $w$ are the components in the local azimuthal, latitudinal and vertical directions, respectively; ${\vec u}_H=\left(u,v,0\right)$ and ${\vec u}_V=\left(0,0,w\right)$ are the horizontal and vertical velocities. Next, we define the fluid buoyancy within the Boussinesq approximation following \cite{Gerkemaetal2008}
\begin{equation}
b=-{\overline g_{\rm eff}\left(z\right)}\frac{\rho'\left(\vec r,t\right)}{{\overline \rho}_{\rm B}},
\end{equation}
where $\rho_{\rm B}$ is a given uniform and constant reference Boussinesq density, and the corresponding Brunt-V\"ais\"al\"a frequency
\begin{eqnarray}
\lefteqn{N^2\left(z\right)=-\frac{\overline g_{\rm eff}}{{\overline \rho}_{\rm B}}\left(\frac{{\rm d}{\overline\rho}}{{\rm d}z}+\frac{{\overline\rho}\,{\overline g}_{\rm eff}}{c_{\rm s}^{2}}\right)}\nonumber\\
&=&-{\overline g_{\rm eff}}\left(\frac{1}{{\overline \rho}_{\rm B}}\,\frac{{\rm d}{\overline\rho}}{{\rm d}z}-\frac{\overline\rho}{{\overline\rho}_{\rm B}}\frac{1}{\Gamma_{1}{\overline P}}\,\frac{{\rm d}{\overline P}}{{\rm d}z}\right),
\end{eqnarray}
where $\rho'$ is the density fluctuation, $\overline\rho$ and $\overline P$ are the reference hydrostatic background density and pressure, respectively, $\Gamma_{1}=\left(\partial\ln{\overline P}/\partial\ln{\overline\rho}\right)_{{\overline S}}$ is the first adiabatic exponent, where ${\overline S}$ is the macroscopic entropy, and $c_{\rm s}=\sqrt{(\Gamma_{1}{\overline P})/{\overline\rho}}$ is the sound speed. In the main text, we will consider that buoyancy is due to the entropy stratification only. However, this is not the case in the oceans and in the interiors of stars and of giant planets where the chemical stratification (the salinity in the case of the oceans) plays an important role \citep[e.g.][]{Thorpe2005,Maeder2009,LeconteChabrier2012}. This more general case is treated in Appendix \ref{Appendix:chemicals}. The three linearised components of the momentum equation are given by:
\begin{equation}
\left\lbrace
\begin{array}{lcl}
\partial_{t}w-{\widetilde f} u=-\displaystyle{\frac{1}{{\overline\rho}_{\rm B}}}\partial_{z}p^{'}+b+\nu\nabla^{2}w\\
\partial_{t}v+f u=-\displaystyle{\frac{1}{{\overline\rho}_{\rm B}}}\partial_{y}p^{'}+\nu\nabla^{2}v\\
\partial_{t}u-f v+{\widetilde f}w=-\displaystyle{\frac{1}{{\overline\rho}_{\rm B}}}\partial_{x}p^{'}+\nu\nabla^{2}u
\end{array}\right.\,,
\label{eq:momentum}
\end{equation}
where $t$ is the time, $p^{'}$ the pressure fluctuation, and $\nu$ the viscosity. Next, we write the continuity equation in the Boussinesq approximation
\begin{equation}
\partial_{z}w+\partial_{y}v+\partial_{x}u=0
\label{eq:continuity}
\end{equation}
and the equation for energy conservation
\begin{equation}
\partial_{t}b+N^{2}\left(z\right) w=\kappa\nabla^{2}w,
\label{eq:energy}
\end{equation}
where $\kappa$ is the thermal diffusivity. Eliminating the horizontal components of the velocity, the pressure and the buoyancy, we reduce the system to the dissipative Poincar\'e equation for the vertical velocity
\begin{equation}
D_{\kappa}D_{\nu}^{2}{\vec\nabla}^{2}w+D_{\kappa}\left(2{\vec\Omega}\cdot{\vec\nabla}\right)^2 w+D_{\nu}\left[N^2 {\nabla}_{\perp}^{2}w\right]=0,
\label{EPD}
\end{equation}
where $D_{\nu}=\left(\partial_{t}-\nu{\vec\nabla}^2\right)$, $D_{\kappa}=\left(\partial_{t}-\kappa{\vec\nabla}^2\right)$ and ${\nabla}_{\perp}^{2}$ is the horizontal Laplacian.

\subsection{Propagation of adiabatic gravito-inertial waves
}\label{subsec:GIW:propagation}

To quantify the transport of momentum by GIWs in our prototype model for any value of $N/2\Omega$, we will follow in \S3, 4 \& 5 the method by \cite{VF2005}, \cite{Zahnetal1997}, \cite{LottGuez2013}, and \cite{Mathis2025}. On the one hand, this means that for the linear damping of GIWs through the viscous and heat diffusion, we will use the so-called quasi-adiabatic method. In this framework, dissipative processes are therefore treated as a first-order perturbation when compared to the adiabatic propagation of waves. On the other hand, we will consider the adiabatic vertical wave number to compute the saturation velocity above which a linear GIW breaks \citep[][]{LottGuez2013,Mathis2025}. As a first step, we thus consider the Poincar\'e equation (Eq. \ref{EPD}) in the limit where $\left\{\nu,\kappa\right\}\rightarrow 0$. We expand the vertical velocity of a monochromatic wave of frequency $\omega$ as 
\begin{equation}
w= W(\chi,z)\exp[\text{i}\omega t],
\label{eq:W}
\end{equation} 
where we recall the definition of the reduced horizontal coordinate $\chi = x \cos\alpha + y \sin\alpha$.  We obtain the adiabatic Poincar\'e equation for GIWs:
\begin{equation}
\left[ N^2(z) - \omega^2 + \tilde{f}_{\text{s}}^2 \right]\partial_{\chi,\chi}W
+ 2f\tilde{f}_{\text{s}}\partial_{z,\chi}W + \left[ f^2 - \omega^2 \right]\partial_{z,z}W  = 0,
\label{eq:EPad}
\end{equation}
where $\tilde{f}_{\text{s}} = \tilde{f}\sin\alpha$. Following  \cite{GS2005}, we introduce the expansion for $\omega\ne f$:
\begin{equation}
W = \hat{W}(z) \exp\left[\text{i}k_{\perp}\left(\chi + \tilde{\delta}z\right)\right],
\label{eq:expansion}
\end{equation}
where
\begin{equation}
\tilde{\delta} = \frac{f\tilde{f}_{\text{s}}}{\omega^2 - f^2}.
\label{eq:delta}
\end{equation}
Substituting Eq. (\ref{eq:expansion}) into Eq. (\ref{eq:EPad}) leads to the equation of a simple harmonic oscillator along the vertical direction:
\begin{equation}
\frac{\text{d}^2 \hat{W}}{\text{d}z^2} + k_V^2\left(z\right) \hat{W} = 0,
\label{eq:Schrodinger}
\end{equation}
where
\begin{equation}
k_V^2 = k_{\perp}^2 \left[ \frac{N^2\left(z\right)-\omega^2}{\omega^2 - f^2} + \left( \frac{\omega \tilde{f}_{\text{s}}}{\omega^2 - f^2}\right)^2\right].
\label{eq:k_z}
\end{equation}
In the case where the GIWs vertical wavelength ($\lambda_{V}=2\pi/k_{V}$) is very small compared to the density height scale ($H_{\rho}$), i.e. $k_V\,H_{\rho}\!>\!\!>\!1$, we can use the Jeffreys-Wentzel-Kramers-Brillouin \citep[hereafter JWKB; we refer the reader to][for a detailed derivation]{FF2005} solutions where ${\widehat W}\left(z\right)\equiv A/k_V^{1/2}\exp[i\,\varepsilon\int_{z_0}^{z} k_V(z'){\rm d}z']$, $A$ being a prescribed amplitude depending on the waves' excitation mechanism. In the case of solar-type stars (where the excitation region is above the propagation zone), we have $\varepsilon=-1$; the phase is propagating outward while the energy is propagating inward in the super-inertial regime. In the case of early-type stars (where the excitation region is below the propagation zone), we have $\varepsilon=1$; the phase is propagating inward while the energy is propagating outward \citep{Mathis2025}. Assuming a uniform local value of the Brunt-V\"ais\"al\"a frequency ($N$), JWKB solutions simplify to plane-wave solutions $\hat{W}(z)={A}^{'}\exp\left(\text{i}\,\varepsilon\,k_V\,z\right)$, where ${A}^{'}$ is the corresponding amplitude. In the case where $k_V^2>0$, we get propagative GIWs along the vertical direction while we have evanescent GIWs in the case where $k_V^2<0$. The vertical component of the velocity becomes in the propagative case:
\begin{equation}
w=A^{'}\exp\left[\text{i}(\tilde{k}_V\,z+k_{\perp}\chi+\omega t) \right],
\label{eq:w}
\end{equation}
where we have introduced the total vertical wave number: 
\begin{equation}
\tilde{k}_V = \varepsilon\,k_V+ k_{\perp}\tilde{\delta}.
\label{eq:total_kz}
\end{equation}
Its non-traditional component $k_{\perp}\tilde{\delta}$ corresponds to the 2D behaviour of the propagation of GIWs when taking the complete Coriolis acceleration into account \citep[we refer the reader to][for a detailed discussion]{DintransRieutord2000,MNT2014}. Indeed, their vertical propagation is coupled to their propagation along the horizontal direction in this case since $\tilde{k}_z$ depends on $ k_{\perp}$. In the polar traditional case, this coupling vanishes since $\delta\rightarrow 0$ and the problem becomes separable.

From a local point of view, at a given position ($z,\theta$), the propagation condition $k_V^2>0$, allows us to identify the possible frequency range for propagative GIWs \citep[][]{GS2005,MNT2014}: 
\begin{equation}
\omega_{-} < \omega < \omega_{+},
\end{equation}
where
\begin{equation}
\omega_{\pm} = \frac{1}{\sqrt{2}} \sqrt{\left[N^2+4\widetilde{\Omega}^2\right] \pm \sqrt{\left[N^2+4\widetilde{\Omega}^2\right]^2 - (2fN)^2}},
\label{eq:freq_spectrum}
\end{equation}
where we have defined a modified rotation rate of the planet
\begin{equation}
\widetilde{\Omega} \equiv \frac{1}{2} \sqrt{f^2 + \tilde{f}_{\text{s}}^2} = \Omega \sqrt{1 - \sin^2\Theta\cos^2\alpha}.
\label{eq:Omega_s}
\end{equation}

From a global point of view, super-inertial waves (with $2\Omega<\omega$) are propagating at all latitudes while they are trapped within an equatorial belt with $\theta\in\left[\theta_{\rm c},\pi-\theta_{\rm c}\right]$ with
\begin{equation}
\theta_c={\rm Arccos}\left[\frac{\omega}{2\Omega}\sqrt{1+\left(\frac{2\Omega}{N}\right)^2\left[1-\left(\frac{\omega}{2\Omega}\right)^2\right]}\,\right]
\end{equation}
in the sub-inertial regime (with $\omega<2\Omega$) \citep[see e.g.][]{Pratetal2016}. We note that in the traditional limit, in which $2\Omega\!<\!\!<\!N$, we recover the classical simplest result $\theta_c={\rm Arccos}\left[\omega/\left(2\Omega\right)\right]$ \citep[e.g.][]{LeeSaio1997}.

\begin{figure*}[t!]
\begin{center}
\includegraphics[width=0.49\textwidth]{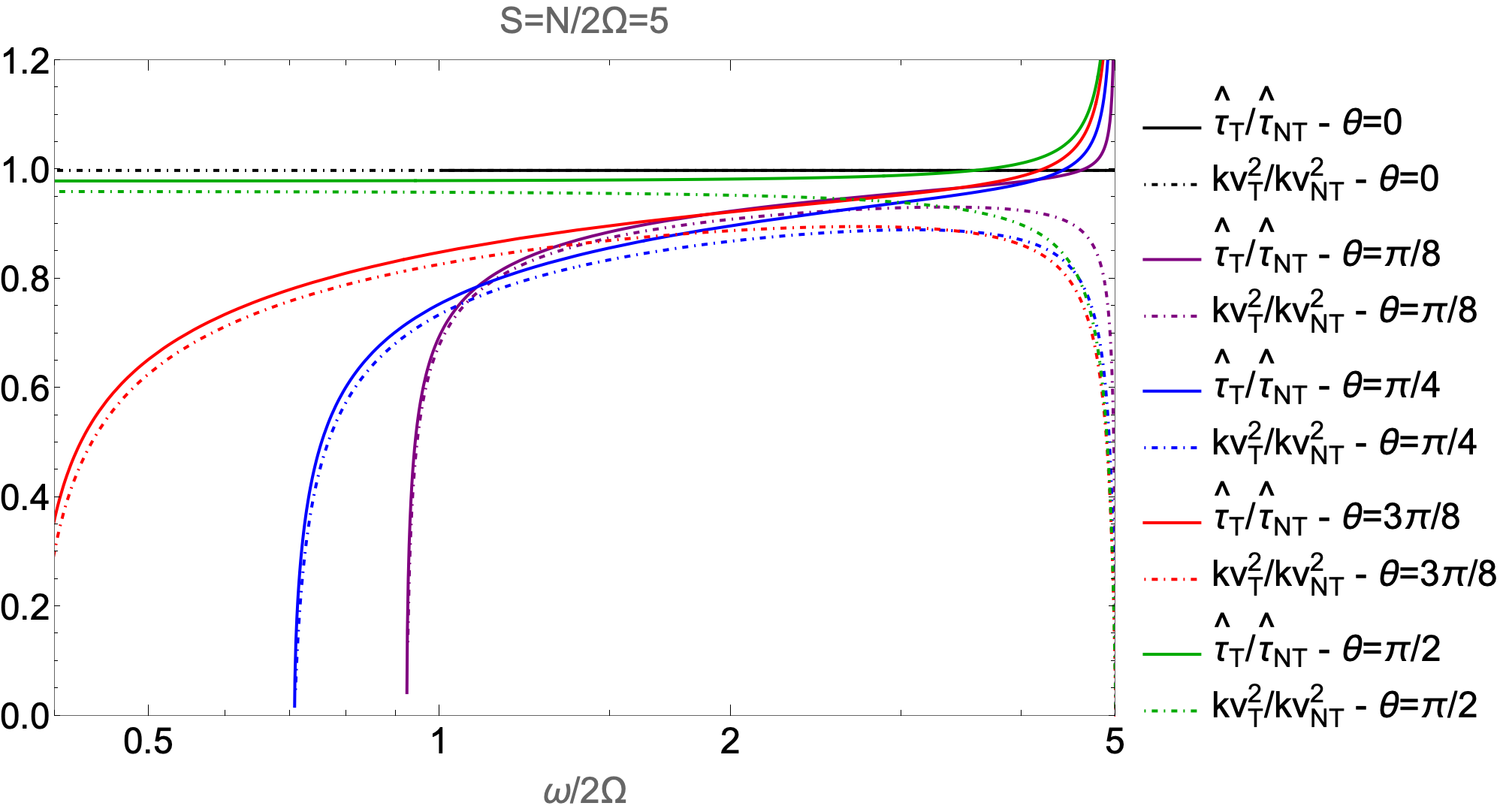}\quad
\includegraphics[width=0.49\textwidth]{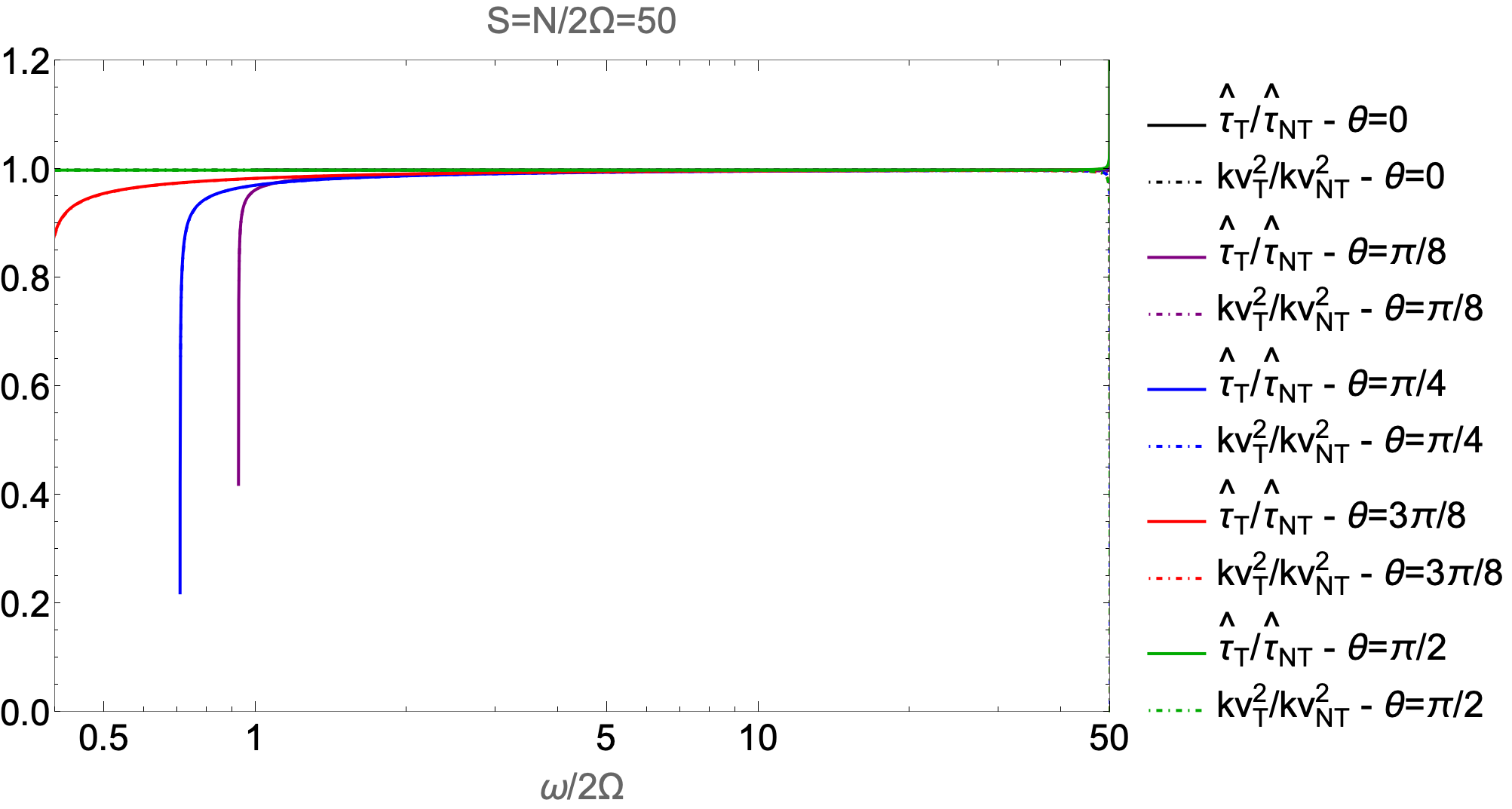}
\end{center}        
\caption{Ratios ${\widehat\tau}_{\rm T}/{\widehat\tau}_{\rm NT}$ and ${k^{2}_{V}}_{\rm T}/{k^{2}_{V}}_{\rm NT}$ as a function of the normalised frequency $\omega/2\Omega$ for different colatitudes $\theta\equiv\left\{0,\pi/8,\pi/4,3\pi/8,\pi/2\right\}$ in the weakly stratified ($N/2\Omega=5$; left panel) and in the strongly stratified ($N/2\Omega=50$; right panel) cases.}
\label{fig:ratiokvldfreq}
\end{figure*}
\subsection{Energy and momentum transport}
\label{subsec:GIW:energy}
The principal objective of our work is to examine the impact of taking into account the full Coriolis acceleration or assuming the TAR in modelling the GIWs - mean zonal flows interactions. Therefore, we introduce the flux of energy transported along the vertical direction by GIWs and the related flux of (angular) momentum. In this section, we thus focus on the energetics of GIWs propagation.\\

In our setup, following \cite{ABM2017}, the energy density flux in the vertical direction is $\left<{p\,w}\right>_{H}$ in real variables, so that we obtain:
\begin{equation}
F_{{\rm E};V}\left(z\right)= \left<{\rm Re}(w){\rm Re}(p)\right>_{H} = \frac{1}{2}\left(\hat{W}\hat{P}^{*}\right),
\label{eq:verticalFluxOfEnergyDef}
\end{equation}
where $^{*}$ is the complex conjugate, ${\rm Re}$ the real part, and  $\left<\cdot\!\cdot\!\cdot\right>_{H}$ the horizontal average. Using the polarisation relation (\ref{eq:polP}), in which $\hat{P}$ is expressed as a function of $\hat{W}$, and considering low-frequency GIWs for which we can assume the JWKB approximation, we get:
\begin{equation}
F_{{\rm E};V} \approx-\frac{1}{2}\,{\varepsilon}\,{\overline\rho}\left(\frac{f^2-\omega^2}{\omega k_{\perp}^2}\right)k_{V}\vert \hat{W} \vert^2.
\label{eq:verticalFluxOfEnergyInTermW}
\end{equation}
In the simplest case of plane waves given in Eq. (\ref{eq:w}), it simplifies into:
\begin{equation}
F_{{\rm E};V} = -\frac{1}{2}\,{\varepsilon}\,{\overline\rho}\left(\frac{f^2-\omega^2}{\omega k_{\perp}^2}\right)k_{V}\vert A^{'}\vert^2,
\label{eq:verticalFluxOfEnergyInTermAmplitudeOfW}
\end{equation}
where we recall that $A^{'}$ is the amplitude of the vertical component of the velocity in this case. In the case of a solar-type (early-type) star, we recover that in the super-inertial regime $F_{{\rm E};V}<0$ (resp. $F_{{\rm E};V}>0$).

We recall that we have introduced in Eq. (\ref{eq:ZFdef}) the mean background zonal flow in an inertial frame:
\begin{equation}
{\vec V}_{\rm 0}=\left({V}_{\Omega}+{\overline U}\left(z\right)\right){\bf e}_{x},
\end{equation}
where $V_{\Omega}=r\sin\theta\,\Omega$ is the zonal flow associated to the global mean rotation $\Omega$ at the point $M$ with spherical coordinates $\left\{r,\theta\right\}$ and ${\overline U}$ is the local horizontally averaged zonal flow. In the momentum equation projected on the local Cartesian basis (Eqs. \ref{eq:momentum}), the presence of this vertically-sheared zonal flow transforms the temporal variation $\partial_t$ into $\partial_t+{\overline U}\partial_x$, where the second term is the advection term due to the mean flow, while the Coriolis acceleration along the longitudinal direction becomes $-f\,v+{\widetilde f}w+w\,\partial_z{\overline U}$. Assuming as a first step that the ratio $R=\partial_z{\overline U}/\left(2\Omega\right)$ is small, the sole modification of the momentum equation is related to the advection term due to the mean flow that leads to the transformation of $\omega$ into the Doppler-shifted frequency ${\widehat\omega}={\omega}+k_x{\overline U}$. Assuming that $R$ is small is close to the weak differential rotation assumption adopted in the stellar case by \cite{Mathisetal2008} to unravel the effects of the global rotation of a stellar radiation zone onto the angular momentum transport by GIWs in a traditional framework.

In this approximation, the evolution equation for the mean zonal flow is given by:   
\begin{equation}
{\overline\rho}\frac{{\rm d}{\overline U}}{{\rm d}t}=\partial_z{F_{{\rm AM};V}}\left(z\right)
\label{eq:MeanZonalFlow}
\end{equation}
with the momentum flux transported by low-frequency GIWs along the vertical direction:
\begin{eqnarray}
\lefteqn{F_{{\rm AM};V}=-\frac{k_x}{{\widehat \omega}}F_{{\rm E};V}=\frac{1}{2}\,{\varepsilon}\,{\overline\rho}\left(\frac{f^2-{\widehat\omega}^2}{{\widehat\omega}^2}\right)\cos\alpha\frac{k_{V}}{k_{\perp}}\vert \hat{W} \vert^2}\nonumber\\
&=&\frac{1}{2}\,{\varepsilon}\,{\overline\rho}\left(\frac{f^2-{\widehat\omega}^2}{{\widehat\omega}^2}\right)\cos\alpha\frac{k_{V}}{k_{\perp}}\vert A^{'}\vert^2,
\label{eq:MomFlux}
\end{eqnarray}
where we have used the relation $k_x=k_{\perp}\cos\alpha$. The prograde (retrograde) waves are such as $k_{x}<0$ (resp. $k_x>0$) and thus $\alpha\in\left[\pi/2,3\pi/2\right]$ (resp. $\alpha\in\left[0,\pi/2\right]{\rm U}\,[3\pi/2,2\pi]$). In the case of super-inertial GIWs propagating in solar-type stars for which $\varepsilon=-1$, we recover that prograde (retrograde) waves deposit (resp. extract) momentum in the radiative core \citep[e.g.][]{Mathisetal2008,Mathis2009}.

\section{Linear dampings}

\subsection{Linear spatial damping rate}
\label{subsec:LSDR}

The first channel that allows GIWs to exchange momentum with mean zonal flows is their linear damping by viscous and heat diffusions. We focus on low-frequency rapidly-oscillating in space GIWs, which are the most damped \citep[e.g.][]{Zahnetal1997,Mathis2009,Alvanetal2015} and which thus transport momentum most efficiently. For these waves, we apply the JWKB approximation. 

In this framework, we write formally the vertical velocity as a quasi-plane wave:
\begin{eqnarray}
\lefteqn{w=E\left(\vec r\right)\exp\left[i\left(\Phi\left(\vec r\right)+\omega t\right)\right]}\nonumber\\
&\equiv& E\left(\vec r\right)\exp\left[i\left(\varepsilon\int_{z_0}^{z}{\widetilde k}_{V}{\rm d}z'+k_{\perp}\chi+\omega t\right)\right],
\label{JWKB}
\end{eqnarray}
where $E$ is an envelope function, which varies slowly compared to the phase $\Phi$, ${\widetilde k}_{V}$ and $k_{\perp}$ are the vertical and horizontal wave numbers, respectively. 

We follow the path of \cite{Zahnetal1997}, where the linear damping is computed considering the local dispersion relation derived from the dissipative wave equation within the Boussinesq approximation. Substituting Eq. (\ref{JWKB}) in Eq. (\ref{EPD}), we obtain the dispersion relation:
\begin{eqnarray}
\lefteqn{\left(\omega-i\kappa k^{2}\right)\left(\omega-i\nu k^{2}\right)^{2}k^{2}=\left(\omega-i\kappa k^{2}\right)\left({\widetilde f}_{s}^{\,\,2}k_{\perp}^{2}+2f{\widetilde f}_{s}\,k_{\perp}{\widetilde k}_{V}+f^{2}\,{\widetilde k}_{V}^{\,2}\right)}\nonumber\\
&+&\left(\omega-i\nu k^2\right)N^2 k_{\perp}^{2}.
\label{M2}
\end{eqnarray}
We assume the quasi-adiabatic approximation, in which we consider the viscous friction and the heat diffusion as first-order perturbations of the adiabatic propagation \citep[see][for stellar interiors and the Earth's atmosphere, respectively]{Press1981,Zahnetal1997,VF2005}. Eq. (\ref{M2}) can be written:
\begin{eqnarray}
    \lefteqn{-ik^2\omega^{-1}\left[\left(\kappa+2\nu\right)\omega^2k^2-\kappa\left(f^2\,{\widetilde k}_V^{\,2}+2f{\widetilde f}_{s}\,{\widetilde k}_Vk_{\perp}+{\widetilde f}_{s}^{\,\,2}k_{\perp}^2\right)-\nu N^2k_{\perp}^2\right]}\nonumber\\
    &=&Ak_{\perp}^2+2Bk_{\perp}{\widetilde k}_V+C{\widetilde k}_V^{\,2},
\label{EqMaitresse}
\end{eqnarray}
where $A=N^2-\omega^2+{\widetilde f}_{s}^{\,\,2}$, $B=f {\widetilde f}_{s}$, and $C=f^2-\omega^2$. 

To derive the spatial damping, we assume that the frequency $\omega$ is real and we expand the vertical wave number as: 
\begin{equation}
{\widetilde k}_V\equiv {\widetilde k}_{V_0}\left(z\right)+{\widetilde k}_{V_1}\left(z\right)={\widetilde k}_{V_0}\left(z\right)+i\,{\widehat\tau}_{j}\left(z\right)
\label{kv}
\end{equation}
with ${\widetilde k}_{V_0}$ the adiabatic vertical wave number and ${\widehat\tau}_{j}$, where $j\equiv\left\{{\rm NT, T},\Omega=0\right\}$ indicates the adopted approximation, the first-order local linear damping rate. Eq. (\ref{EqMaitresse}) is then expanded as:
\begin{eqnarray}
& &-i({\widetilde k}_{V_0}^{\,2}+k_{\perp}^2)\,\omega^{-1}\left[\left(\kappa+2\nu\right)\omega^2({\widetilde k}_{V_0}^{\,2}+k_{\perp}^2)\right.\nonumber\\
& &\left.-\kappa\left(f^2{\widetilde k}_{V_0}^{\,2}+2f {\widetilde f}_{s}\,{\widetilde k}_{V_0}k_{\perp}+{\widetilde f}_{s}^{\,\,2}k_{\perp}^2\right)-\nu N^2k_{\perp}^2\right]\nonumber\\
& &=Ak_{\perp}^2+2Bk_{\perp}{\widetilde k}_{V_0}+i2Bk_{\perp}{\widehat\tau}_{\rm NT}+C{\widetilde k}_{V_0}^{\,2}+i2C{\widetilde k}_{V_0}{\widehat\tau}_{\rm NT}.
\end{eqnarray}
Keeping the zeroth-order terms, we recover the adiabatic dispersion relationship
\begin{equation}
Ak_{\perp}^2+2Bk_{\perp}{\widetilde k}_{V_0}+C{\widetilde k}_{V_0}^{\,2}=0
\end{equation}
while first-order terms allow us to derive the vertical damping rate:
\begin{eqnarray}
\lefteqn{{\widehat\tau}_{\rm NT}=-\frac{1}{2\omega}\frac{{\widetilde k}_{V_0}^{\,2}+k_{\perp}^2}{Bk_{\perp}+C{\widetilde k}_{V_0}}}\nonumber\\
    &&\times\left(\kappa\left[\omega^2({\widetilde k}_{V_0}^{\,2}+k_{\perp}^2)-(f^2{\widetilde k}_{V_0}^{\,2}+2f{\widetilde f}_{s}\,{\widetilde k}_{V_0}k_{\perp}+{\widetilde f}_{s}^{\,\,2}k_{\perp}^2)\right]\right.\nonumber\\
    &&\left.+\nu\left[2\omega^2({\widetilde k}_{V_0}^{\,2}+k_H^2)-N^2k_{\perp}^2\right]\right).
\label{Eq:TauNT}
\end{eqnarray}
When considering a vertically-sheared flow, we can make the substitution $\omega\equiv\widehat\omega$ as soon as the ratio $R=\partial_{z}{\overline U}/\left(2\Omega\right)$ remains small.

The global damping rate as defined by \citep[][]{Press1981,Schatzman1993,Zahnetal1997} is then given by:
\begin{equation}
\tau_{\rm G}\left(z\right)=\varepsilon\int_{z_0}^{z}\vert\,{\widehat\tau_{j}}\,\vert{\rm d}z'.
\end{equation}

\subsection{An example: the stellar case}

\begin{figure*}[h!]
\begin{center}
\includegraphics[width=0.49\textwidth]{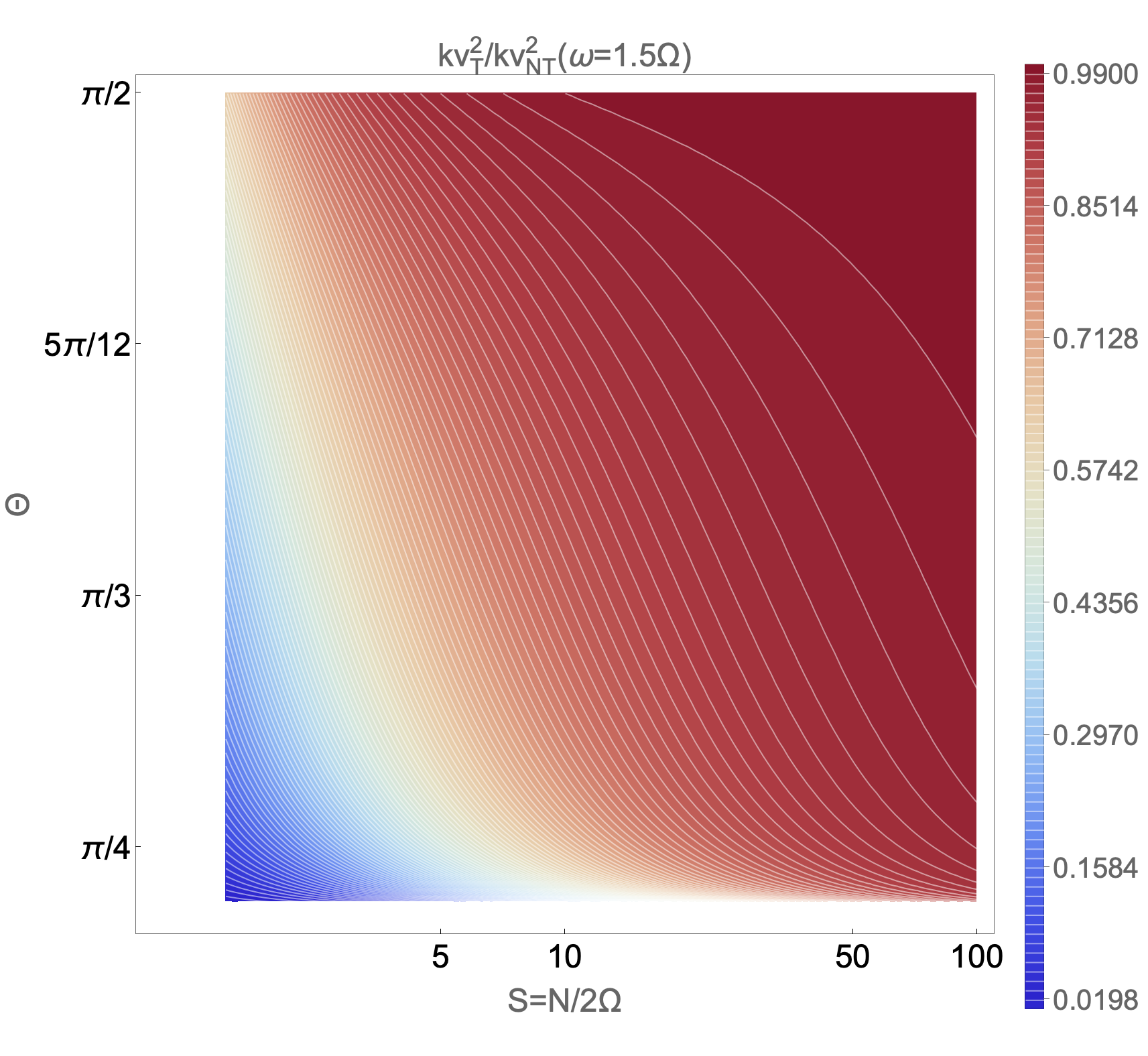}\quad
\includegraphics[width=0.49\textwidth]{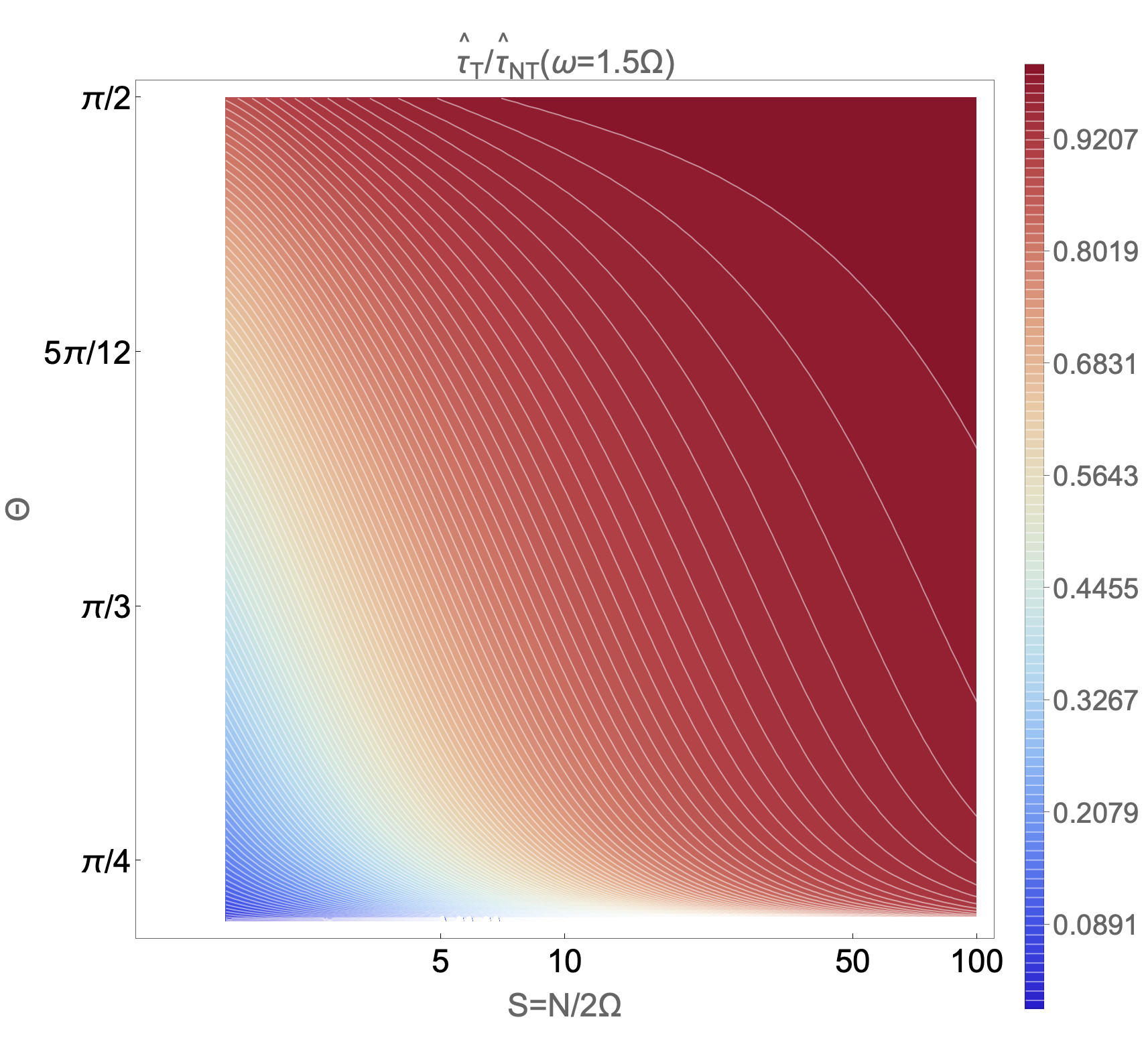}
\includegraphics[width=0.49\textwidth]{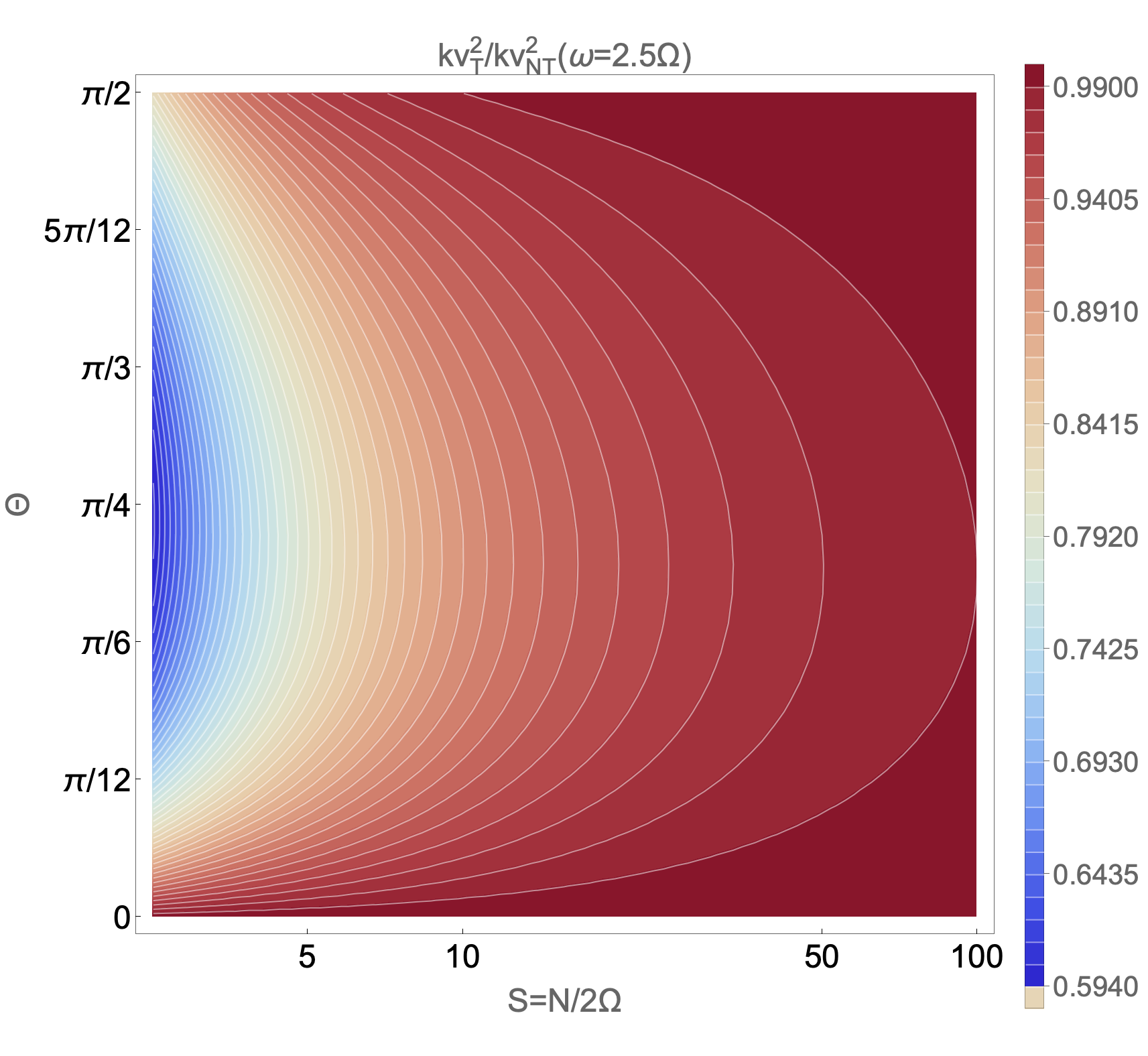}\quad
\includegraphics[width=0.49\textwidth]{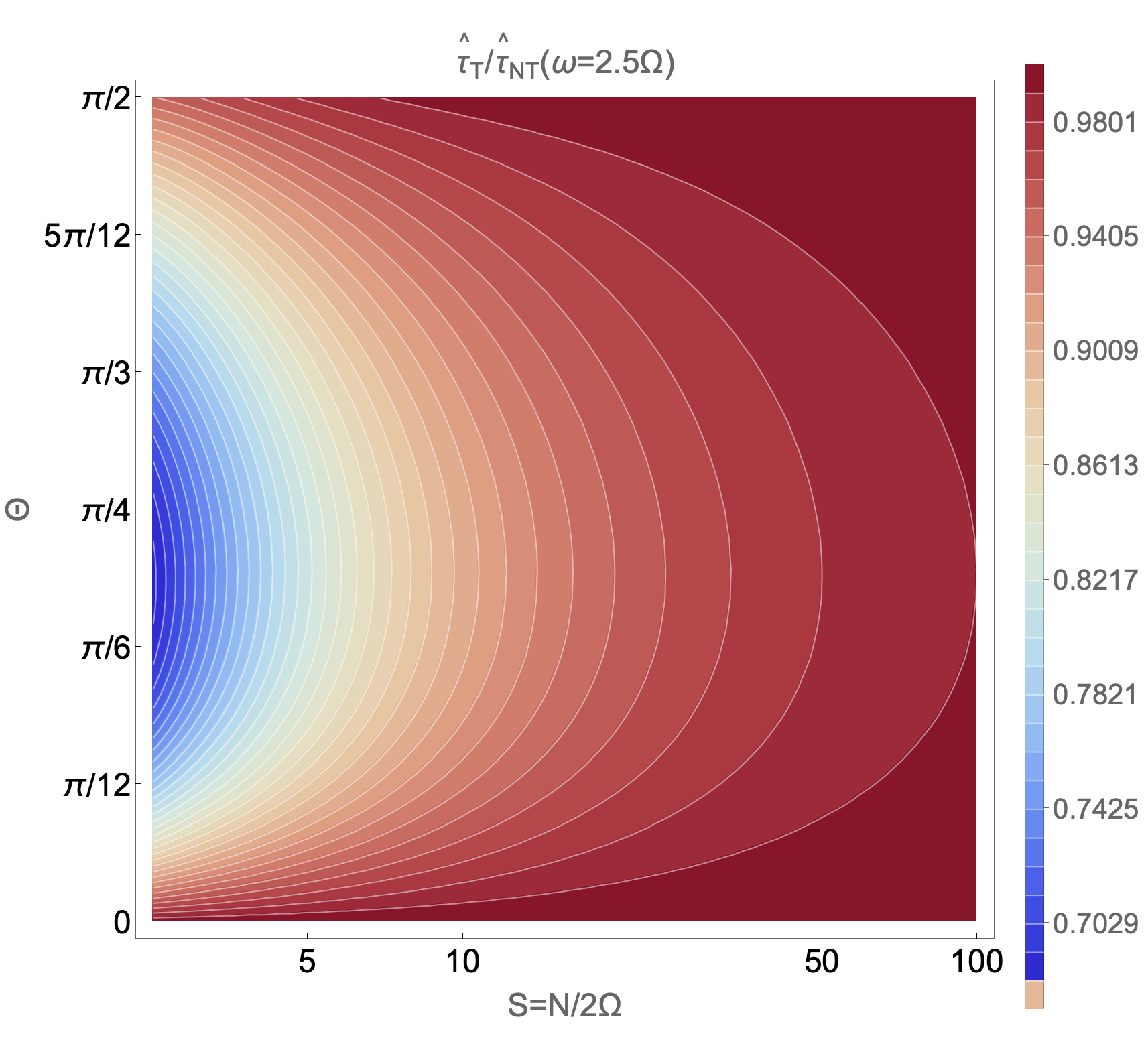}
\end{center}        
\caption{Ratios ${k^{2}_{V}}_{\rm T}/{k^{2}_{V}}_{\rm NT}$ (left column) and ${\widehat\tau}_{\rm T}/{\widehat\tau}_{\rm NT}$ (right column) as a function of the ratio $N/2\Omega$ and of the colatitude $\theta$ in the sub-inertial ($\omega=1.5\Omega$; first line) and in the super-inertial ($\omega=2.5\Omega$; second line) regimes. We recover that in the sub-inertial regime, GIWs are trapped in an equatorial belt while they are propagating at all colatitudes in the super-inertial regime.}
\label{fig:ratiokvldcolat}
\end{figure*}

To demonstrate and illustrate the mandatory use of non-traditional formalisms in the general case and constrain the domain of validity of the TAR as in \cite{ADML2018}, we consider the case of stellar radiation zones where the Prandtl number $P_{r}=\nu/K\!<\!\!<\!1$ (for instance, $P_{r}\approx 10^{-6}$ in the case of the radiative core of the Sun). We therefore focus on the radiative damping as the sole cause of the linear damping of GIWs \citep[e.g.][]{Press1981,Schatzman1993,Zahnetal1997,Mathisetal2008,Mathis2009}. In this limit, the linear non-traditional radiative damping is given by:
\begin{equation}
{\widehat\tau}_{\rm NT}=-\frac{\kappa}{2}\frac{1}{\omega}k_{\perp}^{3}\frac{\left[\left(\delta+\sqrt{\mathcal D}\right)^2+1\right]\left[\omega^{2}\left[\left(\delta+\sqrt{\mathcal D}\right)^2+1\right]-{\mathcal F}\right]}{B+C\left(\delta+\sqrt{\mathcal D}\right)},
\label{tau}
\end{equation}
where
\begin{equation}
\mathcal D=\frac{B^2-AC}{C^2}
\end{equation}
and
\begin{equation}
{\mathcal F}=f^{2}\left(\delta+\sqrt{\mathcal D}\right)^2+2f {\widetilde f}_{s} \left(\delta+\sqrt{\mathcal D}\right) +{\widetilde f}_{s}^{\,\,2}.
\end{equation}
In the traditional case for which ${\widetilde f}={\widetilde f}_{s}=B=\delta=0$, it reduces to 
\begin{equation}
{\widehat\tau}_{\rm T}=\frac{\kappa}{2}k_{\perp}^{3}\frac{N^2\left(N^2-f^2\right)}{\omega\left(\omega^2-f^2\right)^{3/2}\left(N^2-\omega^2\right)^{1/2}}.
\label{tauT}
\end{equation}
In the low-frequency ($\omega\!<\!\!<\!N$) non-rotating case ($f=0$) case, we recover the classical result ${\widehat\tau}_{\Omega=0}=1/2\times\kappa\, k_{\perp}^3\times N^3/{\omega}^4$ \citep[e.g.][]{Press1981,Schatzman1993,Zahnetal1997}.\\

In Fig. \ref{fig:ratiokvldfreq}, we study the variation of the ratios of the vertical linear damping rate and of the adiabatic vertical wave number computed assuming the TAR by those computed in the non-traditional case as a function of the wave frequency normalised by the inertial frequency ($\omega/2\Omega$) in a weakly stratified case for which $S=N/2\Omega=5$ (left) and in a strongly stratified one for which $S=N/2\Omega=50$ (right) for different co-latitudes $\theta=\left\{0,\pi/8,\pi/4,3\pi/8,\pi/2\right\}$. We choose the specific case $\alpha=\pi/2$ (i.e. the axisymmetric GIWs for which $k_x=0$) to maximise the non-traditional effects since ${\tilde f}_{s}={\tilde f}\sin\alpha$. We see that except at the pole (i.e. in the traditional case) and at the equator, these ratios are decreasing when diminishing the normalised frequency ($\omega/2\Omega$). This decreasing is becoming stronger when entering in the sub-inertial regime (i.e. $\omega<2\Omega$) and tends to $\infty$ with $\left\{{{k^{2}_{V}}_{\rm T}}/{k^{2}_{V}}_{\rm NT},{\widehat\tau}_{\rm T}/{\widehat\tau}_{\rm NT}\right\}\rightarrow 0$ when $\omega\rightarrow\omega_{-}$. This means that the TAR can strongly underestimate the vertical wave number, and as a consequence the linear damping, in the sub-inertial regime. This demonstrates the necessity to adopt the non-traditional framework when studying the transport of momentum and matter by GIWs. This is also true in the strongly stratified regime in which however the non-traditional prediction converges very rapidly towards the traditional one in the super-inertial regime in which $N/(2\Omega)\!>\!\!>\!1$ while $\omega\!<\!\!<\!N$.\\

\noindent In Fig. \ref{fig:ratiokvldcolat}, we illustrate these predictions with a complementary point of view. We plot the ratios ${{k^{2}_{V}}_{\rm T}}/{k^{2}_{V}}_{\rm NT}$ (left raw) and ${\widehat\tau}_{\rm T}/{\widehat\tau}_{\rm NT}$ (right raw) as a function of the ratio $N/\left(2\Omega\right)$ and of the colatitude $\theta$ for a sub-inertial wave for which $\omega=1.5\Omega$ (first line) and a super-inertial wave for which $\omega=2.5\Omega$ (second line). On the one hand, we verify again that the non-traditional predictions converge to the traditional ones as expected in the strongly-stratified regime in which $N/(2\Omega)\!>\!\!>\!1$ while $\omega\!<\!\!<\!N$ \citep[e.g.][]{Friedlander1987,Mathisetal2008,Mathis2009,MNT2014}. On the other hand, we finds again the strong underestimation of the vertical wave number, and as a consequence of the linear damping in the sub-inertial regime as this was shown in Fig. \ref{fig:ratiokvldfreq}. In the sub-inertial regime, the strongest underestimation occurs for weak $N/(2\Omega)$ ratios close to the critical co-latitude $\theta_c$. This can be related to the so-called waves' focussing appearing in the non-traditional case where GIWs become strongly focused and sheared close to $\theta_c$. This can lead to strong mixing in weakly stratified oceanic layers \citep[][]{ShriraTownsend2010}. In the super-inertial regime, the strongest underestimation (with moderate values $\sim0.6-0.7$) occurs for weak $N/(2\Omega)$ ratios at mid-latitudes.          

\section{Wave breaking}

\subsection{Convective wave breaking}

\subsubsection{Formalism}

We now consider the second mechanism through which GIWs exchange momentum with mean flows: their convective breaking (hereafter CWB). To derive the transported flux of momentum, we follow the method as proposed in \cite{LottGuez2013} for non-orographic gravity waves and generalised to GIWs by \cite{Ribsteinetal2022} in the traditional case \citep[see also][in the case of a global deep spherical shell]{Mathis2025}. It is important to underline that the predictions obtained within these theoretical frameworks have been successfully compared to in-situ measurments of waves-triggered Reynolds stresses in the stratosphere \citep[][]{Lottetal2023}. This demonstrates simultaneously the realism and the robustness of the proposed parametrisation and the relevance of the method allowing its derivation. One can note that the conditions in the Earth's atmosphere, where the Prandtl number $P_r$ is close to unity ($P_r\approx 0.71$), strongly differ from those of stellar radiation zones in which $P_r\!<\!\!<\!1$. However, since the adiabatic heat equation (Eq. \ref{eq:heatCWB}) used to derive the convective breaking condition does not depend on $P_r$, we can hope that this approach can also be applied in stellar interiors but also in oceanic layers in which $P_r$ is above unity ($P_r\approx 7$).\\

\begin{figure*}[h!]
\begin{center}
\includegraphics[width=0.49\textwidth]{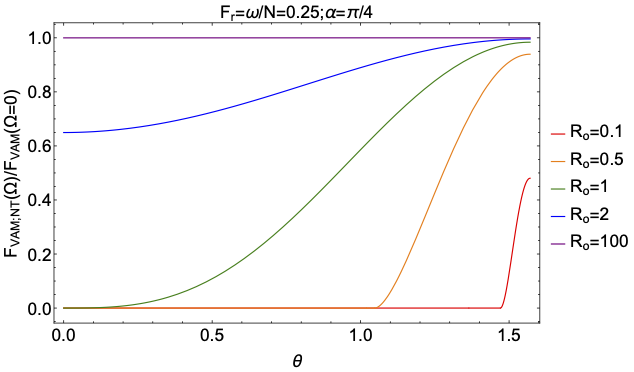}\quad
\includegraphics[width=0.49\textwidth]{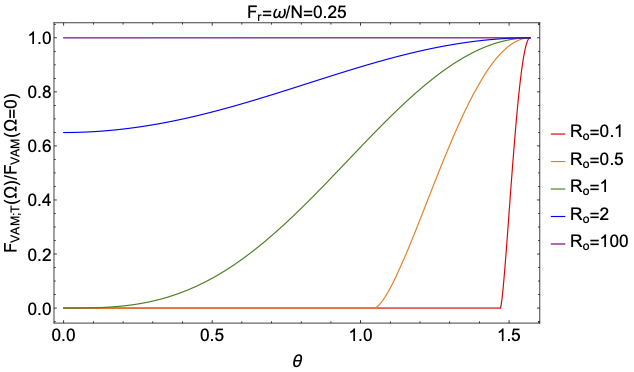}
\end{center}        
\caption{Ratio of the vertical flux of momentum computed with taking into account the full Coriolis acceleration (left panel) and assuming the TAR (right panel) with its value computed in the non-rotating case (with $\Omega\equiv0$) as a function of the colatitude ($\theta$) for a fixed value of the wave's Froude number ($F_r=\omega/N=0.25$) and different values of the wave's Rossby number ($R_o=\left\{0.1,0.5,1,2,100\right\}$). In the non-traditional case, we have fixed $\alpha=\pi/4$.}
\label{fig:RatioFluxCWB}
\end{figure*}

As a first step, we consider the heat transport equation expressed as a function of the fluctuation of the temperature ($T^{'}$):
\begin{equation}
D_{t}T^{'}+\Gamma w=0\quad\hbox{with}\quad\Gamma=\frac{{\overline T}N^{2}}{\overline g},
\label{eq:heatCWB}
\end{equation}
where $D_{t}=\partial_{t}+{\overline U}\partial_{x}$. Following \cite{Lindzen1981}, \cite{LottGuez2013}, \cite{Ribsteinetal2022}, and \cite{Mathis2025}, the condition for the wave convective breaking writes:
\begin{equation}
\vert\partial_{z}T^{'}\vert>\Gamma.
\end{equation}
Using the Fourier expansion in time in the linearised heat transport equation (\ref{eq:heatCWB}) and considering low-frequency GIWs for which the JWKB approximation can be used, we derive the saturation amplitude of the vertical velocity for which $\vert\partial_{z}T^{'}\vert=\Gamma$:  
\begin{equation}
\vert {\widehat W} \vert_{\rm sat}^{\rm CWB}=\frac{{\widehat \omega}}{k_V}
\label{eq:satWCWB}
\end{equation}
above which GIWs are subject to convective breaking. We obtain a result similar in its mathematical form to both \cite{LottGuez2013} and \cite{Mathis2025}. Moreover, as this has been already pointed out by \cite{Mathis2025}, this result is equivalent to breaking criteria proposed by \cite{Press1981}, \cite{GoodmanDickson1998} and \cite{BarkerOgilvie2010}. This can be explained by the fact that the Coriolis acceleration does not intervene directly in the heat transport equation; its action will be through the modification of the vertical wave number $k_V$. This allows us, using Eq. (\ref{eq:verticalFluxOfEnergyInTermW}), to compute the carried flux of energy along the vertical direction:
\begin{equation}
F_{\rm E}^{\rm CWB}=-\frac{1}{2}\,\varepsilon\,{\overline \rho}\frac{\left(f^2-{\widehat \omega}^{2}\right)}{{\widehat\omega}\,k_{\perp}^{2}}k_{V}\left(\frac{{\widehat \omega}}{k_V}\right)^2.
\end{equation}
Finally, we derive the flux of momentum along this direction using Eq. (\ref{eq:verticalFluxOfEnergyInTermAmplitudeOfW}): 
\begin{eqnarray}
\lefteqn{F_{\rm AM}^{\rm CWB}=-\frac{k_x}{{\widehat \omega}}F_{\rm E}}\nonumber\\
&=&\frac{1}{2}\,\varepsilon\,{\overline\rho}\left(\frac{f^2-{\widehat\omega}^2}{{\widehat\omega}^2}\right)\cos\alpha\frac{k_{\rm V}}{k_{\perp}}\left(\frac{{\widehat\omega}}{k_V}\right)^2,
\end{eqnarray}
where $k_x=k_{\perp}\cos\alpha$. Introducing
\begin{eqnarray}
&&F_r=\frac{\widehat\omega}{N}, R_{o}=\frac{\widehat\omega}{2\Omega},\nonumber\\ 
&&R_{o;V}=\frac{{\widehat\omega}}{f}=\frac{1}{\cos\theta}R_{o},\,\,\hbox{and}\,\,R_{o;H}=\frac{{\widehat \omega}}{\widetilde f}=\frac{1}{\sin\theta}R_{o}
\end{eqnarray}
we can writte 
\begin{eqnarray}
\lefteqn{F_{\rm AM}^{\rm CWB}=\frac{1}{2}\,{\varepsilon}\,{\overline\rho}\left(R_{o;V}^{-2}-1\right)\cos\alpha}\nonumber\\
&\times&\left[\frac{F_{r}^{-2}-1}{1-R_{o;V}^{-2}}+\frac{R_{o;H}^{-2}\sin^2\alpha}{\left(1-R_{o;V}^{-2}\right)^2}\right]^{-1/2}\left(\frac{\widehat\omega}{k_{\perp}}\right)^2;
\end{eqnarray}
sub-inertial (respectively super-inertial) waves correspond to $R_{o}<1$ (respectively $R_{o}>1$).

In the non-rotating case, this reduces to the result obtained by \cite{LottGuez2013}:
\begin{equation}
\vert F_{\rm AM}^{\rm CWB}\left(\Omega=0\right)\vert=\frac{1}{2}{\overline\rho}\cos\alpha\,F_{r}\,\left(\frac{\widehat\omega}{k_{\perp}}\right)^2.
\end{equation}
Using the expression for the flux of momentum transported along the vertical direction given in Eq. (\ref{eq:MomFlux}), we can also compute explicitly the ratio of its value in the non-traditional case by the one in the traditional case:
\begin{equation}
\frac{F_{\rm AM;NT}}{F_{\rm AM;T}}=\frac{{k_{V}}_{\rm NT}}{{k_{V}}_{\rm T}}\left(\frac{\vert {\widehat W}_{\rm NT} \vert}{\vert {\widehat W}_{\rm T}\vert}\right)^2.
\label{eq:RatioFluxMom}
\end{equation}
Using the saturation amplitude obtained in Eq. (\ref{eq:satWCWB}), we get
\begin{equation}
\frac{F_{\rm AM;NT}^{\rm CWB}}{F_{\rm AM;T}^{\rm CWB}}=\left(\frac{{k_V}_{\rm T}}{{k_V}_{\rm NT}}\right)<1.
\label{eq:RatioFluxCWBNTT}
\end{equation}

\subsubsection{Exploration of the parameter space}

\begin{figure}[h!]
\begin{center}
\includegraphics[width=0.49\textwidth]{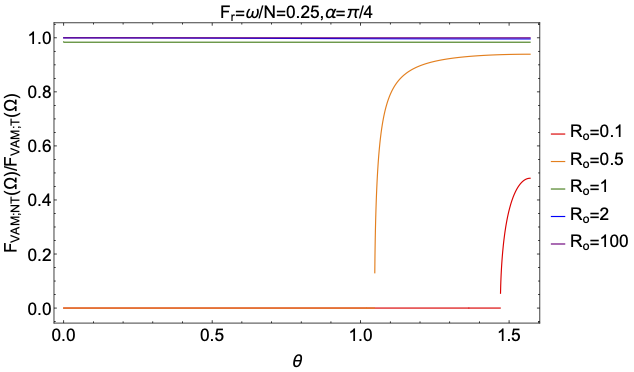}\quad
\end{center}        
\caption{Ratio of the vertical flux of momentum computed with taking into account the full Coriolis acceleration with its value computed in the traditional case as a function of the colatitude ($\theta$) for a fixed value of the wave's Froude number ($F_r=\omega/N=0.25$) and different values of the wave's Rossby number ($R_o=\left\{0.1,0.5,1,2,100\right\}$). In the non-traditional case, we have fixed $\alpha=\pi/4$.}
\label{fig:RatioFluxCWBNTT}
\end{figure}

In Fig. \ref{fig:RatioFluxCWB}, we examine the ratio of the flux of momentum carried at the convective breaking by GIWs with the one carried by gravity waves (corresponding to the non-rotating case $\Omega\equiv 0$) as a function of the co-latitude $\theta$ for fixed wave Froude ($F_{\rm r}={\widehat\omega}/N$) and Rossby ($R_{\rm o}={\widehat\omega}/\left(2\Omega\right)$) numbers in the non-traditional (left panel) and in the traditional (right panel) cases. In both cases, we see that for decreasing waves Rossby number (i.e. for increasing rotation at fixed frequency) these ratio are decreasing. This can be interpreted keeping in mind how rotation inhibits the convective instability and the triggered transport of energy and as a consequence of momentum \citep{Stevenson1979,AM2019}. In the sub-inertial regime (i.e. for $R_{\rm o}<1$), the flux of angular momentum vanishes outside an equatorial belt for $\theta<\theta_{c}$ and $\theta>\pi-\theta_{c}$. This corresponds to the GIWs equatorial trapping by the Coriolis acceleration in the sub-inertial regime \citep[e.g.][]{DintransRieutord2000,GS2005,MNT2014}. The key difference between the traditional and the non-traditional regime is the artificial convergence of the ratio to unity at the equator predicted by the TAR. This convergence is artificial since the TAR cannot be applied at the equator because the projection of the rotation vector along the vertical direction vanishes there. The correct prediction is therefore the one obtained in the non-traditional framework taking into account the full Coriolis acceleration. In this regime, we see that the carried flux of momentum diminishes at all the co-latitudes when compared to the non-rotating case, in particular at the equator. This difference between the non-traditional and the traditional cases is evaluated in Fig. \ref{fig:RatioFluxCWBNTT} where we plot the ratio of the flux predicted in the non-traditional regime and of the one predicted in the traditional regime as predicted in Eq. (\ref{eq:RatioFluxCWBNTT}). In the sub-inertial regime, it is well bellow unity since ${k_V}_{\rm T}<{k_V}_{\rm NT}$. This can be understood since taking into account the full Coriolis acceleration inhibits more efficiently the convective instability of the waves and the related transport. In the super-inertial regime, the ratio converges to unity. This is coherent with the results obtained in the previous section studying linear radiative damping, where the non-traditional prediction converges to the traditional one when $R_{\rm 0}\rightarrow\infty$. As a general conclusion, we therefore see how it is mandatory to use the non-traditional framework to study wave-mean flow interactions in particular in the equatorial regions. On the one hand, those regions are of particular importance in planetary atmospheres where these interactions play a key role in the general circulation, for instance with the Quasi-Biennal Oscillation \citep[e.g.][and references therein]{Baldwinetal2001}. On the other hand, in the case of rapidly-rotating stars, sub-inertial GIWs trapped in the equatorial belt efficiently deposit angular momentum below the stellar surface that can lead to mass decretion \citep{Ando1985,Rogers2015,Neineretal2020}.    

\subsection{The shear-induced GIWs breaking}

As shown by \cite{GLS1991} and \cite{Mathis2025}, internal gravity waves and GIWs are not only subject to convective breaking. They can become unstable to their own vertical shear, that can lead to their breaking. Hereafter, we will note SWB for this Shear-driven Wave Breaking. Following \cite{Mathis2025}, we adapt the methodology built by \cite{Lottetal2012} and \cite{LottGuez2013} for the study of the convective wave breaking to the case of the SWB. To do so, we consider the instability criteria for the vertical shear of GIWs from which we derive an estimate of the saturation amplitude for the vertical component of the velocity and the corresponding flux of momentum transported along the vertical direction.\\ 

Before deriving such quantities it is important to report the progresses that have been achieved recently on the study of the instability of a vertical shear in a stably stratified rotating fluid with taking into account the full Coriolis acceleration. In a recent theoretical article, \cite{ParkMathis2025} have demonstrated that such a vertical shear is submitted to two different types of instabilities: the inflectional instability that corresponds to the classical vertical shear instability studied in stellar physics \citep[e.g.][]{Zahn1983,Zahn1992,Maeder1996,PratL2013,PratL2014,GK2016} and the inertial instabilities such as the instability of Rayleigh unstable differential rotation profiles and the Goldreich-Schubert-Fricke instability in stellar physics \citep[e.g.][]{GS1967,Fricke1968,Barkeretal2020,Dymottetal2023}. \cite{ParkMathis2025} demonstrated that the inflectional instability develops mostly around the equator for non-axisymmetric perturbations while baroclinic instabilities develop at mid-latitudes for stably-stratified rotating fluids with a weak heat diffusion and around the poles for stably-stratified rotating fluids with a strong heat diffusion as in the case of stellar radiation zones. In this work, we are focusing on the GIWs-zonal flows interactions and this particularly in the sub-inertial regime in which GIWs are trapped in a belt around the equator. We thus choose to focus on the vertical-shear inflectional instability while vertical-shear inertial instabilities would be important for the potential breaking of Rossby waves that can propagate at high latitudes. One of the key result by \cite{ParkMathis2025} is that the theoretical inflectional instability criteria around the equator derived within a linear stability analysis is close to the one for the non-rotating case. This has also been observed in direct numerical simulations around the equator of vertical shear instabilities by \cite{ChangGaraud2021}. We will thus consider the following instability criteria:    
\begin{equation}
\frac{N^2}{\left|\displaystyle{\frac{{\rm d}}{{\rm d}z}\left[\sqrt{\left<{\rm Re}\left({\vec u}_{H}\right)\cdot{\rm Re}\left({\vec u}_{H}\right)\right>_{H}}\right]}\right|^2}<Ri_{\rm c}\quad\hbox{and}\,
\label{Richardson1}
\end{equation}
\begin{equation}
\frac{N^2}{\left|\displaystyle{\frac{{\rm d}}{{\rm d}z}\left[\sqrt{\left<{\rm Re}\left({\vec u}_{H}\right)\cdot{\rm Re}\left({\vec u}_{H}\right)\right>_{H}}\right]}\right|^2}\left(\frac{v\,l}{K}\right)<Ri_{\rm c}
\label{Richardson2}
\end{equation}
in the cases of stably-stratified rotating fluids that are the seat of weak and strong heat diffusion, respectively. We recall that $\left<\cdot\!\cdot\!\cdot\right>_{H}$ is the horizontal average with $\left<{\rm Re}\left(\vec X\right)\cdot{\rm Re}\left(\vec X\right)\right>_{H}=1/2\left({\vec X}\cdot{\vec X}^{*}\right)$ and we introduce $u_H=\sqrt{{\vec u}_{H}\cdot{\vec u}_{H}^{*}}$ the amplitude of the horizontal component of GIWs and $v$ and $l$ the characteristic vertical velocity and length scale that must be considered in the regime of strong heat diffusion. As in \cite{Mathis2025}, we identify that $v\equiv\,\sqrt{\left<{\rm Re}\left({\vec u}_{V}\right)\cdot{\rm Re}\left({\vec u}_{V}\right)\right>_{H}}$, the amplitude of the vertical component of the studied GIW, and $l\equiv 2\pi\,k_{V}^{-1}$, its vertical wave number. In the forthcoming sections, each of these regimes will therefore be studied.

\subsubsection{The regime of weak heat diffusion}
This first regime corresponds to the geophysical case where the heat diffusion does not dominate the viscous one (we recall that the Prandtl number, $P_{r}$, is of the order of unity in the Earth's atmosphere while it is above unity in the oceans with $P_{r}\sim 7$). Considering low-frequency GIWs for which the JWKB approximation can be used, the Richardson criteria given in Eq. (\ref{Richardson1}) becomes:
\begin{equation}
\frac{N^2}{1/2\,k_{V}^2\left|u_{H}\right|^2}<Ri_{\rm c}.
\end{equation}
Following the methodology presented in \cite{Mathis2025}, we express the horizontal average of the horizontal component of GIWs using polarization relationships Eqs. (\ref{eq:polU}) and (\ref{eq:polV}) and the JWKB approximation. We obtain:
\begin{equation}
\left|u_{H}\right|^2=\left|\widehat U\right|^2+\left|\widehat V\right|^2=\left(\frac{k_{V}}{k_{\perp}}\right)^2\frac{f^2+{\widehat\omega}^2}{{\widehat\omega}^2}\left|{\widehat W}\right|^2.
\label{eq:uH}
\end{equation}
The threshold instability condition allows us to derive the expression for the saturation amplitude of the vertical velocity of GIWs in the case where their vertical shear triggers their breaking:
\begin{equation}
\left(\left|{\widehat W}\right|^2\right)_{\rm sat}^{\rm SWB}=\frac{2}{Ri_{\rm c}}\frac{N^2}{f^2+{\widehat\omega}^2}\left(\frac{k_{\perp}}{k_{V}}\right)^2\left(\frac{\widehat\omega}{k_{V}}\right)^2.
\label{eq:SatWeakDiff}
\end{equation}
In the non-rotating case, we verify that we recover the result obtained by \cite{Mathis2025} in the simplest case of internal gravity waves:
\begin{equation}
\left(\left|{\widehat W}\right|^2\right)_{\rm sat}^{\rm SWB}\left(\Omega=0\right)=\frac{2}{Ri_{\rm c}}\left(\frac{\widehat\omega}{k_{V}}\right)^2.
\end{equation}
Using again the expression for the flux of momentum transported along the vertical direction given in Eq. (\ref{eq:MomFlux}), we can compute the ratio of its value in the non-traditional case by the one in the traditional case:
\begin{equation}
\frac{F_{\rm AM;NT}}{F_{\rm AM;T}}=\frac{{k_{V}}_{\rm NT}}{{k_{V}}_{\rm T}}\left(\frac{\vert {\widehat W}_{\rm NT} \vert}{\vert {\widehat W}_{\rm T}\vert}\right)^2.
\label{eq:RatioFluxMom}
\end{equation}
Using the saturation amplitude obtained in Eq. (\ref{eq:SatWeakDiff}), we get
\begin{equation}
\frac{F_{\rm AM;NT}^{\rm SWB}}{F_{\rm AM;T}^{\rm SWB}}=\left(\frac{{k_V}_{\rm T}}{{k_V}_{\rm NT}}\right)^3<1.
\end{equation}
As in the case of the convective wave breaking, we therefore obtain a predicted flux when taking the full Coriolis acceleration into account weaker than the one predicted within the TAR. Since ${k_V}_{\rm T}<{k_V}_{\rm NT}$, the shear instability can be favoured in the non-traditional case but with a saturation amplitude for the vertical velocity weaker than in the traditional prediction that leads to a weaker flux of momentum. The change in power of ${{k_V}_{\rm T}}/{{k_V}_{\rm NT}}$ of the ratio $F_{\rm AM;NT}^{\rm SWB}/F_{\rm AM;T}^{\rm SWB}$ from unity obtained in the case of the CWB to $3$ is due to the modulation factor $N^2/(f^2+{\widehat \omega}^2)\,\left(k_{\perp}/k_{V}\right)^2$ in Eq. (\ref{eq:SatWeakDiff}). 

\subsubsection{The regime of strong heat diffusion}

This regime corresponds to the case of the bulk of stellar radiative regions where the Prandtl number is small. We again consider low-frequency GIWs for which the JWKB approximation can be applied. Following the same methodology that in the weak heat diffusion regime, the vertical shear instability criteria (\ref{Richardson2}) becomes: 
\begin{equation}
\frac{N^2}{1/2\,k_{V}^2\left|u_{H}\right|^2}\frac{1/2\,\left|{\widehat W}\right|^2\,2\pi\,k_{V}^{-1}}{K}<Ri_{\rm c}.
\end{equation}
Using the polarisation equation (\ref{eq:uH}), it leads to the saturation amplitude for the vertical component of the velocity: 
\begin{equation}
\left|{\widehat W}\right|_{\rm sat}^{\rm SWB}=\frac{2\pi}{Ri_{\rm c}}\frac{N^2}{f^2+{\widehat\omega}^2}\left(\frac{k_{\perp}}{k_{V}}\right)^2\frac{\widehat\omega}{K\,k_{V}^2}\frac{\widehat\omega}{k_{V}}.
\end{equation}
Using Eq. (\ref{eq:RatioFluxMom}), we finally obtain the ratio of the value of the flux of momentum transported along the vertical direction predicted in the non-traditional case by the one in the traditional case:
\begin{equation}
\frac{F_{\rm AM;NT}^{\rm SWB}}{F_{\rm AM;T}^{\rm SWB}}=\left(\frac{{k_{V}}_{\rm T}}{{k_{V}}_{\rm NT}}\right)^9<1.
\end{equation}
Its overestimation in the traditional regime is therefore more severe in the case of the shear-driven breaking of GIWs in the strong heat diffusion stellar regime than in their shear-driven breaking in the weak heat diffusion atmospheric or oceanic regimes and in the case of their convective breaking.

\section{Wave transport parametrisation}

In this work, we have demonstrated that the full Coriolis acceleration must be taken into account in the general case where the conditions for validity of the TAR are not satisfied. Indeed, if we apply this approximation outside of its range of applicability we will underestimate the linear damping of GIWs in the sub-inertial regime and predict a momentum deposition at an altitude where it has been already damped in reality, while we will overestimate the transported flux of momentum by GIWs convective and shear-driven breakings. As in the case ignoring rotation \citep[][]{LottGuez2013} and in the traditional case \citep[][]{Ribsteinetal2022}, we thus have to provide a parametrisation of momentum transport by GIWs. Following \cite{LottGuez2013}, we thus derive the transported flux of momentum:
\begin{eqnarray}
\lefteqn{F_{\rm AM}\left(z\right)=-\frac{k_x}{{\widehat\omega}}F_{\rm E}\left(z\right)}\nonumber\\
&&=\sum_{\omega,k_{\perp}}{\rm min}\left\{F_{\rm AM}\left(z=z_{\rm o}\right)\exp\left[-{\tau}_{\rm G}\left(z\right)\right],\right.\nonumber\\
&&{\left.F_{\rm AM;NT}^{\rm CWB}\left(z\right),F_{\rm AM;NT}^{\rm SWB}\left(z\right)\right\}}.
\end{eqnarray}
It is given by the minimum value between the classical quasi-adiabatic flux that applies during the GIWs propagation before the altitude of the breaking and the saturated values of the flux at this place and after. 

Let us explain how the derived formalism is working. We consider a given altitude $z$. On the one hand, if the quasi-adiabatic velocity of a GIW is smaller than the breaking saturation velocities (i.e. before the altitude where the wave is breaking), the corresponding quasi-adiabatic flux of momentum is smaller than the one carried by the breaking of the wave and the min function therefore selects the quasi-adiabatic value. On the other hand, if the quasi-adiabatic velocity of a GIW has formally become larger than one of the breaking saturation velocities (i.e. after the wave has broken), the quasi-adiabatic flux becomes formally larger than the one carried by the breaking, which is naturally selected by the min function.

This allows us to compute the variation of the mean zonal flow $\overline U$ varying with the altitude following Eq. (\ref{eq:MeanZonalFlow}).\\

In our formalism, we assume that GIWs can be convectively unstable and that their own vertical shear can be unstable \citep[][]{Thorpe1999,Sutherland2001}. The formalism is based on local linear stability criteria and selects the instability with the lowest threshold velocity leading to one of the instabilities and related breaking. However, the literature show us that: i) internal gravity waves and GIWs can be submitted to other types of instabilities such as parametric subharmonic instabilities \citep[e.g.][]{GM2019}, ii) the trigger of the different instabilities depends on the properties of the initial perturbations \citep[e.g.][]{LombardRiley1996,Liuetal2010}, and iii) several waves' instabilities (for instance the convective and the shear-induced ones) can take place in a flow simultaneously \citep[e.g.][]{LombardRiley1996,Howlandetal2021}, all these mechanisms being related to subtle wave-wave nonlinear interactions. It shows how devoted nonlinear numerical simulations must be undertaken to better characterize these complex nonlinear mechanisms and improve their parametrization for long time-scale evolution \citep[e.g.][]{Frumanetal2014,Gervaisetal2021}.

\section{Discussion and conclusions}

In this work, we have built a prototype local Cartesian model to study the interactions between gravito-inertial (inertia-gravity) waves and stellar (planetary/geophysical) mean flows. It allows us to treat any stratification, mean rotation rate, heat diffusion and viscosity. We consider two of the key mechanisms allowing waves to exchange momentum with mean flows: their linear dampings by viscosity and heat diffusion \citep[e.g.][]{Press1981,Zahnetal1997,VF2005} and their nonlinear breaking because of their convective \citep[][]{Lindzen1981,Lottetal2012,LottGuez2013} and shear \citep[e.g.][]{GLS1991,Mathis2025} instabilities. Our local model can be considered as a first predictive tool of what happens in regions in stellar, planetary, atmospheric, and oceanic layers close to critical layers where the linear dampings of the waves drastically increases until they nonlinearly break. In the case of the linear damping, we demonstrate that it increases when we compare the rotating case to the non-rotating case as already predicted in the literature \citep[e.g.][]{Pantillonetal2007,Mathisetal2008}. This is due to the increase of the wave number. The key new interesting result, obtained here in the case where the heat diffusion is dominating (i.e. in the case of stellar radiation zones), is that the linear damping predicted using the traditional approximation is completely underestimated in the sub-inertial regime when compared to the non-traditional prediction in which the full Coriolis acceleration is taken into account. This means that in the case where the buoyancy associated to the stable stratification does not dominate the component of the Coriolis acceleration in the direction of the entropy/chemical stratification, we must abandon the traditional approximation which would predict a deposit of momentum far away from the excitation region of waves than it is in reality. This result is of great importance when studying transport and mixing in weakly stably stratified rotating stellar radiative regions in rapidly rotating young late-type stars and early-type stars. The same conclusion applies for weakly stratified atmospheric and oceanic layers where the damping will be driven by the combination of heat diffusion and viscosity or dominated by viscosity, respectively \citep[note that a similar conclusion has been obtained recently by][]{Slepyshev2025}. When considering the waves' convective and shear-induced nonlinear breaking, we find that the saturation velocity at which the transition to breaking occurs decreases because of the increase of the wave number. This leads to a decrease of the efficiency of the momentum transport. For the convective wave breaking, this can be explained because of the loss of efficiency of convection in rapidly rotating systems \citep[e.g.][]{Stevenson1979,Barkeretal2014,AM2019,Currieetal2020}. For the shear-induced breaking, the inflexional instability of a vertically-sheared flow is localised around the equator with an instability criteria similar to the non-rotating case \citep{ParkMathis2025}. The loss of efficiency of the induced momentum transport in this case is then due to the fact that the non-traditional vertical wave vector is larger than the traditional one outside the equator and the poles. Once again, we notice a strong difference between the predictions obtained using the traditional approximation or going beyond it. The traditional approximation overestimates the momentum transport because it underestimates the vertical wave number that increases because of the non-traditional terms. This result is of great importance again for rapidly rotating stars where gravito-inertial waves are propagating while such waves are key players to drive and regulate the atmospheric and oceanic general circulation on our Earth \citep[e.g.][]{MacKinnonetal2017,Sutherlandetal2019,Ribsteinetal2022,Achatzetal2024}. Key perspectives of this work are now to combine this complete non-traditional modelling to a corresponding wave excitation model \citep{MNT2014,Augustsonetal2020}, to be able to generalise it to global geometry, and to take into account magnetic fields \citep[e.g.][]{RogersMacGregor2011,MdB2012}. This would be key developments for stellar structure and secular evolution models and for atmospheric/oceanic general circulation models in which transport of momentum and macroscopic mixing of chemicals have to be parametrised in a robust and tractable way.

\begin{acknowledgements}
S.M. warmly thanks the referee, Pr. A. Barker, for his constructive and detailed report, which allows him to improve the initial manuscript. S.M. acknowledges support from the European Research Council (ERC) under the Horizon Europe programme (Synergy Grant agreement 101071505: 4D-STAR), from the CNES SOHO-GOLF and PLATO grants at CEA-DAp, and from PNPS (CNRS/INSU). While partially funded by the European Union, views and opinions expressed are however those of the author only and do not necessarily reflect those of the European Union or the European Research Council. Neither the European Union nor the granting authority can be held responsible for them. This work has been written within the time of the Vend\'ee Globe 2024 race.
\end{acknowledgements}

\bibliographystyle{aa}  
\bibliography{Mathis2025} 

\begin{appendix}

\section{Polarization relations}

\begin{figure*}[h!]
\begin{center}
\includegraphics[width=0.49\textwidth]{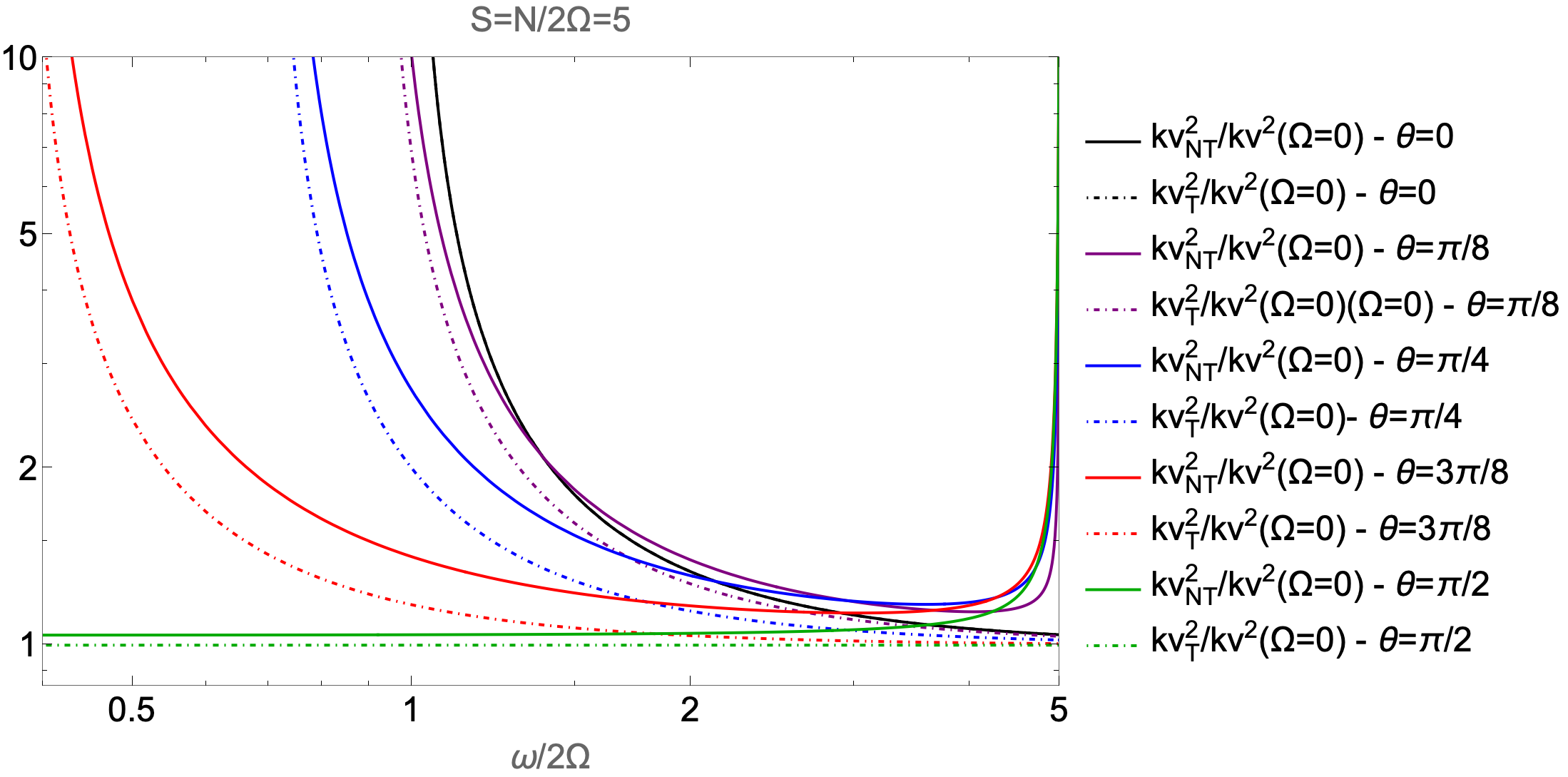}\quad
\includegraphics[width=0.49\textwidth]{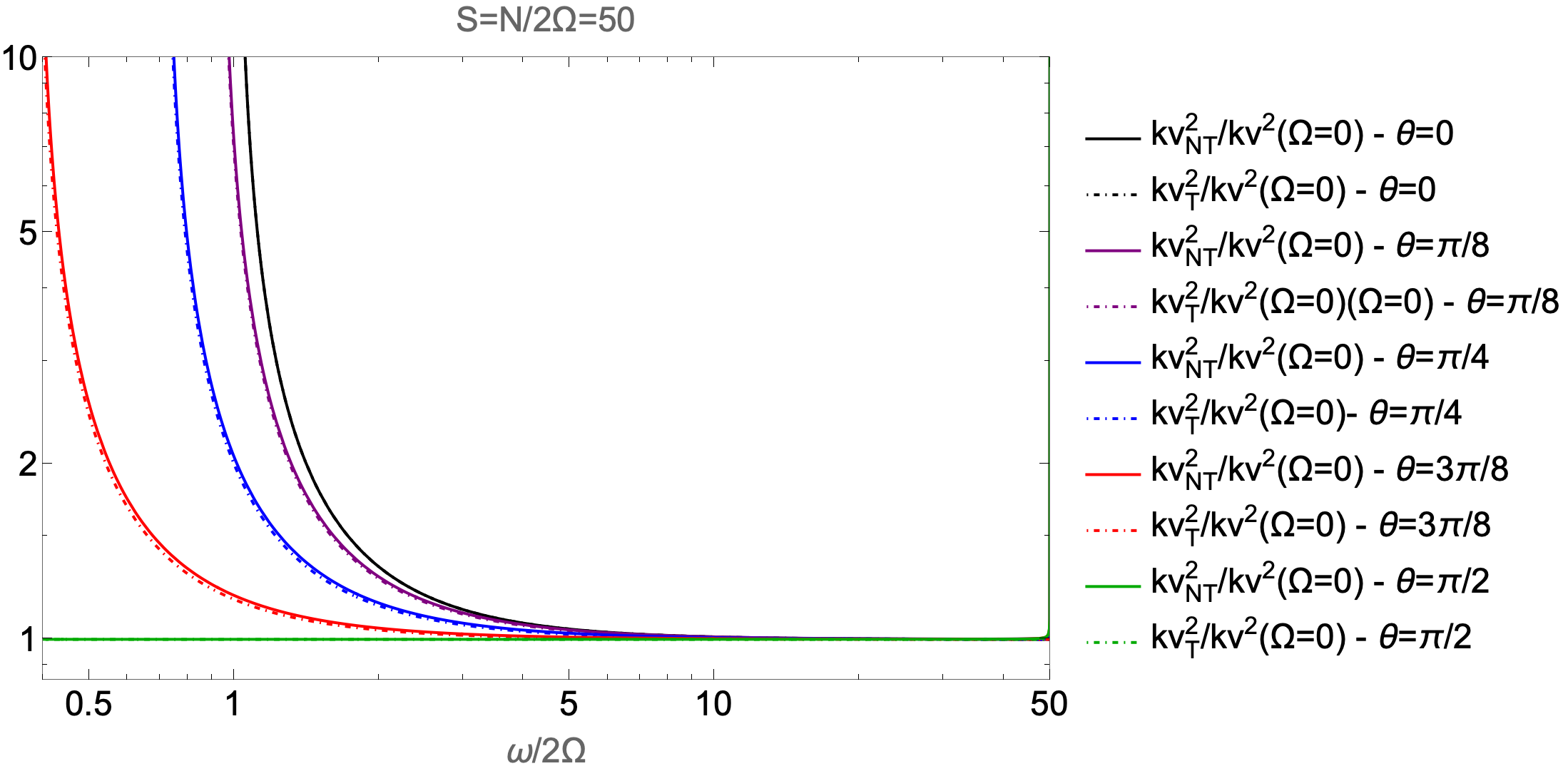}
\end{center}        
\caption{Ratios ${k_V^2}_{\rm NT}/{k_V^2}\left(\Omega=0\right)$ and ${k_V^2}_{\rm T}/{k_V^2}\left(\Omega=0\right)$ as a function of the normalised frequency $\omega/2\Omega$ for different colatitudes $\theta\equiv\left\{0,\pi/8,\pi/4,3\pi/8,\pi/2\right\}$ in the weakly stratified ($N/2\Omega=5$; left panel) and in the strongly stratified ($N/2\Omega=50$; right panel) cases.}
\label{Fig:Appen1}
\end{figure*}

\begin{figure*}[h!]
\begin{center}
\includegraphics[width=0.49\textwidth]{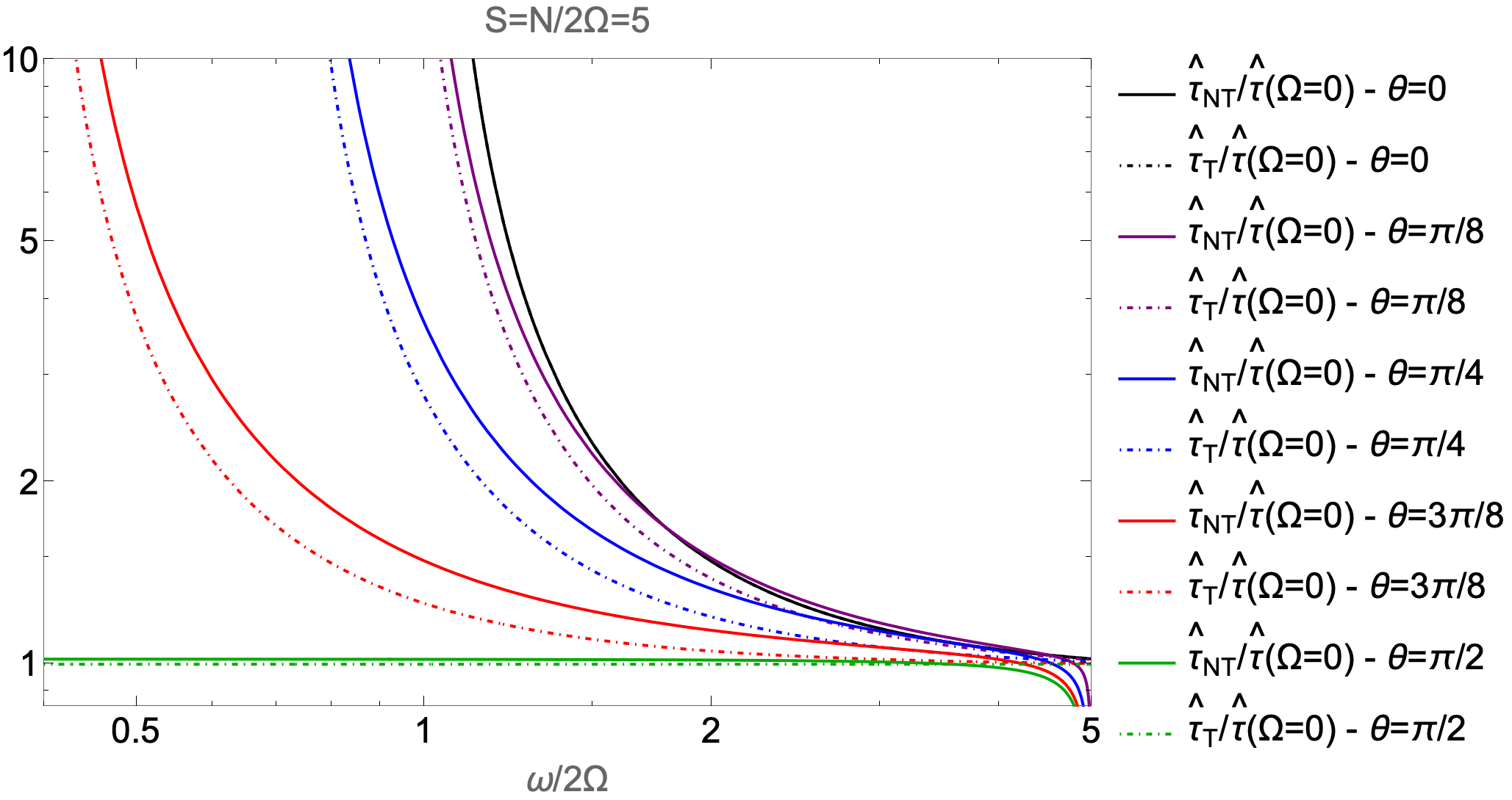}\quad
\includegraphics[width=0.49\textwidth]{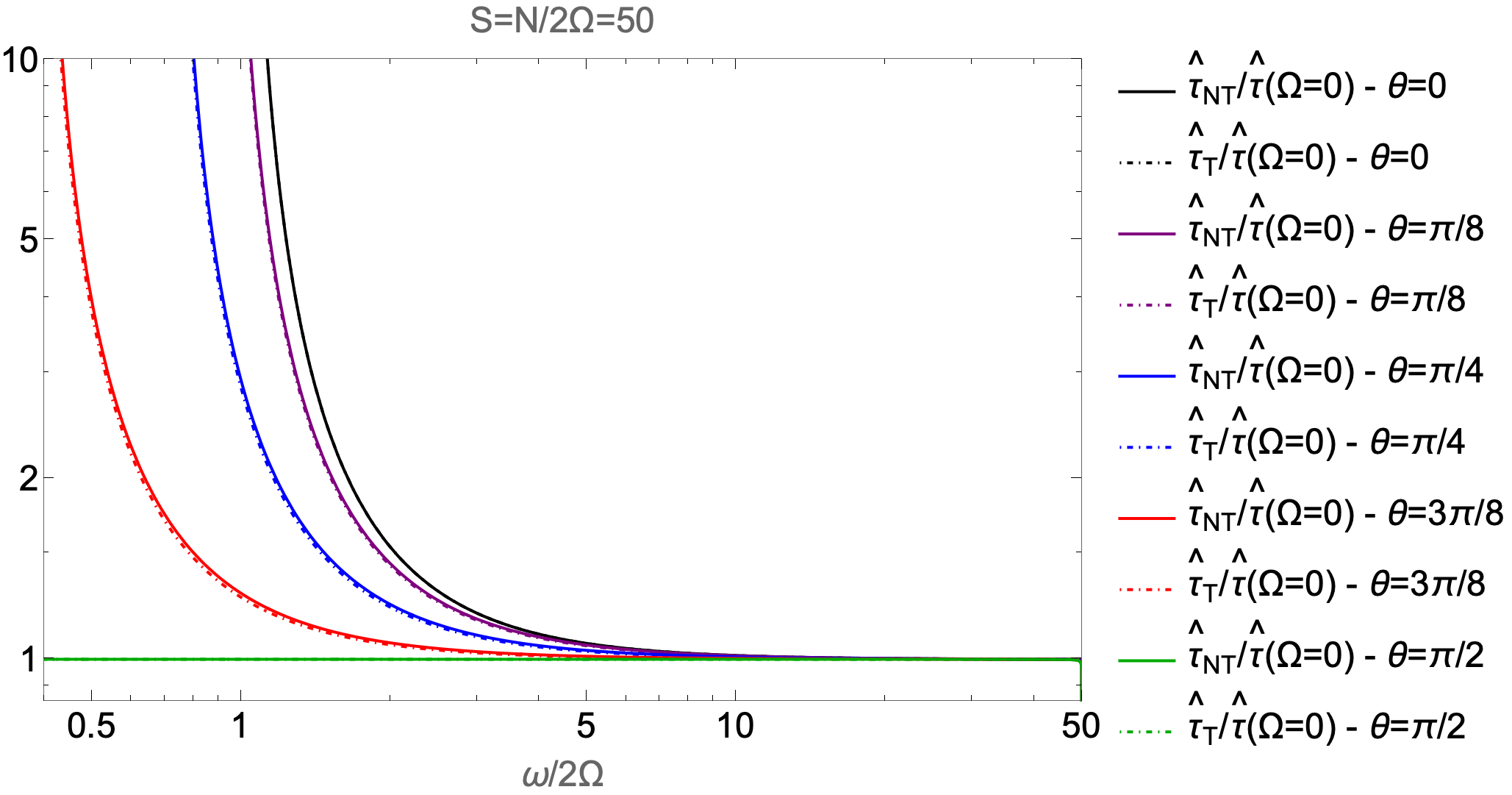}
\end{center}        
\caption{Ratios ${\widehat\tau}_{\rm NT}/{\widehat\tau}\left(\Omega=0\right)$ and ${\widehat\tau}_{\rm T}/{\widehat\tau}\left(\Omega=0\right)$ as a function of the normalised frequency $\omega/2\Omega$ for different colatitudes $\theta\equiv\left\{0,\pi/8,\pi/4,3\pi/8,\pi/2\right\}$ in the weakly stratified ($N/2\Omega=5$; left panel) and in the strongly stratified ($N/2\Omega=50$; right panel) cases.}
\label{Fig:Appen2}
\end{figure*}

The main goal of our so-called prototype local model is to quantify the GIWs - zonal mean flow interactions and the related flux of energy and momentum these waves are transporting. To reach this objective, we need to compute the product of the vertical velocity of GIWs by their fluctuation of pressure. To do so in an efficient way, the best method is to derive the expression of the horizontal components of the velocity, the fluctuation of pressure and the buoyancy as a function of the vertical component of the velocity through the so-called polarization relations.\\ 

To derive these latters, we follow \cite{ADML2018} and we express each component of the velocity and each perturbation of a thermodynamical quantity as this has been done in Eqs. (\ref{eq:W},\ref{eq:expansion}) for the vertical velocity: 
\begin{equation}
{\widetilde x}\left(\vec r,t\right) = {\rm Re}\left\{X(\chi,z)\text{e}^{\text{i}\omega t}\right\}\quad\hbox{where}\quad X = \hat{X}(z) \exp\left[\text{i}k_{\perp}(\chi + \tilde{\delta}z)\right]
\label{eq:transformation_allfields}
\end{equation}
and ${\rm Re}$ is the real part of a complex number. Since in the non-traditional case, the Poincar\'e equation is not separable, $\hat{X}$ carries only part of the vertical dependence of the field ${\widetilde x}$, the other part being included in $\exp\left(\text{i}k_{\perp}\tilde{\delta}z\right)$. A separable solution will be $X=\hat{X}(z)\exp\left(\text{i}k_{\perp}\chi\right)$ without any dependence of the vertical function on $k_{\perp}$. In the case of a global model going beyond the local model presented here, the term $\exp\left(\text{i}k_{\perp}\tilde{\delta}z\right)$ corresponds to the mixed derivative $\partial_{z,\chi}$ in Eq. (\ref{EPD}), which is responsible for the non-separability of the Poincar\'e equation in the general case.\\ 

The linearised momentum equation Eq (\ref{eq:momentum}) becomes in its inviscid limit:
\begin{align}
& \text{i}\omega U - fV + \tilde{f}W = -\frac{\text{i}k_{\perp}\cos\alpha}{\overline\rho}P,\\
& \text{i}\omega V + fU = -\frac{\text{i}k_{\perp}\sin\alpha}{\overline\rho} P,\\
& \text{i}\omega W - \tilde{f}U = -\frac{1}{\overline\rho}\frac{\partial P}{\partial z} + B,
\end{align}
while we obtain for the equations of conservation of mass (Eq. \ref{eq:continuity}) and energy in its adiabatic limit (Eq. \ref{eq:continuity}), respectively:
\begin{gather}
\text{i}k_{\perp}\cos\alpha U + \text{i}k_{\perp}\sin\alpha V + \frac{\partial W}{\partial z} = 0,\\[2mm]
\text{i}\omega B + N^2 W = 0.
\end{gather}
Using Eq. (\ref{eq:transformation_allfields}), we have:
\begin{equation*}
\frac{\partial W}{\partial z} = \left(\hat{W}' + \text{i}k_{\perp}\tilde{\delta}\,\hat{W}\right)\exp\left[\text{i}k_{\perp}(\chi + \tilde{\delta}z)\right],
\end{equation*}
where the prime denotes differentiation with respect to $z$, and all the fields can be expressed in term of $\hat{W}$ as :
\begin{align}
\hat{U} &= \frac{\tilde{f}_{\text{s}}(f\cos\alpha - \text{i}\omega\sin\alpha)}{f^2-\omega^2}\hat{W} + \frac{f\sin\alpha + \text{i}\omega\cos\alpha}{\omega k_{\perp}}\hat{W}', \label{eq:polU}\\
\hat{V} &= \frac{\tilde{f}_{\text{s}}(f\sin\alpha + \text{i}\omega\cos\alpha)}{f^2-\omega^2}\hat{W} - \frac{f\cos\alpha - \text{i}\omega\sin\alpha}{\omega k_{\perp}}\hat{W}',\label{eq:polV}\\
\hat{P} &= \text{i}\frac{\rho_0\tilde{f}_{\text{c}}}{k_{\perp}}\hat{W} + \text{i}\frac{\rho_0(f^2-\omega^2)}{\omega k_{\perp}^2}\hat{W}',\label{eq:polP}\\
\hat{B} &= \text{i}\frac{N^2}{\omega}\hat{W},
\label{eq:polB}
\end{align}
where $\tilde{f}_{\text{c}}=\tilde{f}\cos\alpha$. Then, each field has to be multiplied by $\exp \left[\text{i}\left(k_{\perp}(\chi + \tilde{\delta}z) + \omega t\right)\right]$ to get the complete solution. The physical solution is obtained by taking the real part of the complete complex solution.\\
In the case of low-frequency GIWs, for which the JWKB assumption can be assumed, we thus obtain:
\begin{equation}
\hat{P} \approx -{\varepsilon}\,{\overline\rho}\frac{\left(f^2-\omega^2\right)}{\omega\,k_{\perp}^2}\,k_V\,\hat{W}.
\end{equation}

\section{Comparison with the non-rotating regime}

In the main text, we present and discuss the ratios ${k_V^2}_{\rm T}/{k_V^2}_{\rm NT}$ and ${{\widehat\tau}}_{\rm T}/{{\widehat\tau}}_{\rm NT}$ to identify when it is necessary to take into account the full Coriolis acceleration. It is also interesting to show as a complement in Figs. \ref{Fig:Appen1} and \ref{Fig:Appen2} the ratio ${k_V^2}_{\rm NT}/{k_V^2}\left(\Omega=0\right)$, ${k_V^2}_{\rm T}/{k_V^2}\left(\Omega=0\right)$, ${\widehat\tau}_{\rm NT}/{\widehat\tau}\left(\Omega=0\right)$ and ${\widehat\tau}_{\rm T}/{\widehat\tau}\left(\Omega=0\right)$ respectively in the weakly (left panels) and strongly (right panels) stratified cases in which $S=N/2\Omega=5$ and $50$, respectively. This allows us to identify how both $k_V^2$ and ${\widehat\tau}$ are increasing at low-frequencies with rotation. This leads to linear dampings closer to the excitation region and to transported vertical flux of momentum by convective and shear-induced wave breakings weaker in the rotating case than in the non-rotating case (we refer the reader to \S3 \& 4, respectively). These trends are predicted both when assuming the TAR and in a fully non-traditional framework. Finally, we also recover that ${k^{2}_{V}}_{\rm T}/{k^{2}_{V}}_{\rm NT}<1$ and ${\widehat\tau}_{\rm T}/{\widehat\tau}_{\rm NT}<1$.

\section{General case with both entropy and chemical stratification}
\label{Appendix:chemicals}

\subsection{Linear damping}

In the case where we have to take into account both the entropy (temperature) and chemicals stratification, we have to introduce the general equation of state:
\begin{equation}
\frac{\partial\rho}{\rho}=\alpha\frac{\partial P}{P}-\delta\frac{\partial T}{T}+\phi\frac{\partial\mu}{\mu},
\end{equation}
where
\begin{equation}
\alpha=\left(\frac{\partial\ln\rho}{\partial\ln P}\right)_{T,\mu},\,\delta=-\left(\frac{\partial\ln\rho}{\partial\ln T}\right)_{P,\mu}, \phi=\left(\frac{\partial\ln\rho}{\partial\ln\mu}\right)_{P,T},
\end{equation}
and $\mu$ is the molecular weight in the stellar and the planetary case \citep{KW1994} and the salinity in the oceanic case \citep{Thorpe2005}. Following \cite{Dhouibetal2024}, the transport equation for the chemical composition is written:
\begin{equation}
D_{t}\mu=\lambda_{\mu}{\nabla}^{2}{\mu},
\end{equation}
where $D_{t}=\partial_{t}+\left({\vec v}\cdot\nabla\right)$ and $\lambda_{\mu}$ is the chemical diffusion assumed to be constant.
We recall the macroscopic entropy transport equation:
\begin{equation}
{\rho}T\,D_{t}S=\rho\,c_{P}\,\kappa{\nabla}^{2}T,
\end{equation}
where $c_{P}$ is the specific heat capacity, and the relation expressing the entropy as a function of the temperature and the pressure:
\begin{equation}
{\rm d}S=c_{P}\left(\frac{{\rm d}T}{T}-\nabla_{\rm ad}\frac{{\rm d}P}{P}\right),
\end{equation}
where $\nabla_{\rm ad}=\left(\partial \ln T/\partial \ln P\right)_{S}$.
 
Linearising those two equations and combining them with the momentum and continuity equation we get the Poincar\'e equation for the vertical velocity \citep{Worthemetal1983}:
\begin{eqnarray}
\lefteqn{D_{\kappa}D_{\lambda_{\mu}}D_{\nu}^{2}{\vec\nabla}^{2}w+D_{\kappa}D_{\lambda_{\mu}}\left(2{\vec\Omega}\cdot{\vec\nabla}\right)^2 w+D_{\nu}D_{\lambda_{\mu}}\left[N_{T}^2 {\nabla}_{\perp}^{2}w\right]}\nonumber\\
&&+D_{\nu}D_{\kappa}\left[N_{\mu}^2 {\nabla}_{\perp}^{2}w\right]=0,
\label{EPDC}
\end{eqnarray}
where $D_{\lambda_{\mu}}=\left(\partial_{t}-\lambda_{\mu}{\nabla}^{2}\right)$. We have introduced the thermal and chemical Brunt-V\"ais\"al\"a frequencies: 
\begin{equation}
N_{T}^{2}=\frac{{\overline g}\delta}{H_{P}}\left(\nabla_{\rm ad}-\nabla\right)\quad\hbox{and}\quad N_{\mu}^{2}=\frac{{\overline g}\phi}{H_{P}}\nabla_{\mu}
\end{equation}
with ${\overline g}$ the gravity, $\nabla=\left(\partial\ln{\overline T}/\partial\ln{\overline P}\right)$ and $\nabla_{\mu}=\left(\partial\ln{\overline \mu}/\partial\ln{\overline P}\right)$ and we define the total Brunt-V\"ais\"al\"a frequency:
\begin{equation}
N^2=N_{T}^{2}+N_{\mu}^{2}.
\end{equation}
Following the same method as in \S \ref{subsec:LSDR}, where the linear spatial damping rate is derived adopting a local Boussinesq quasi-adiabatic modeling, we derive the quasi-adiabatic dispersion relation 
\begin{equation}
 \begin{split} 
&-ik^2\omega^{-1}\times\bigl[\left(\kappa+\lambda_{\mu}+2\nu\right)\omega^2k^2\\
&-\left(\kappa+\lambda_{\mu}\right)\left(f^2\,{\widetilde k}_V^{\,2}+2f{\widetilde f}_{s}\,{\widetilde k}_Vk_{\perp}+{\widetilde f}_{s}^{\,\,2}k_{\perp}^2\right)-\left(\nu N^2 + \kappa N_{\mu}^{2} + \lambda_{\mu} N_{T}^{2}\right)k_{\perp}^2\bigr]\\
&=Ak_{\perp}^2+2Bk_{\perp}{\widetilde k}_V+C{\widetilde k}_V^{\,2}.
\end{split}
\label{EqMaitresseC}
\end{equation}
We assume that the frequency $\omega$ is real and we expand the vertical wave number as: 
\begin{equation}
{\widetilde k}_V\equiv {\widetilde k}_{V_0}\left(z\right)+{\widetilde k}_{V_1}\left(z\right)={\widetilde k}_{V_0}\left(z\right)+i\,{\widehat\tau}_{j}\left(z\right),
\end{equation}
where ${\widetilde k}_{V_0}$ is the adiabatic vertical wave number while ${\widehat\tau}_{j}$, with $j\equiv\left\{{\rm NT, T},\Omega=0\right\}$ indicating the adopted approximation, is the first-order local linear damping rate given by: 
\begin{equation}
{\widehat\tau}_{\rm NT}=-\frac{1}{2\omega}\frac{{\widetilde k}_{V_0}^{\,2}+k_{\perp}^2}{Bk_{\perp}+C{\widetilde k}_{V_0}}\left(\kappa\,D+\lambda_{\mu}\,E+\nu\,F\right),
 \end{equation}
 with
\begin{equation}
D=\omega^2({\widetilde k}_{V_0}^{\,2}+k_{\perp}^2)-(f^2{\widetilde k}_{V_0}^{\,2}+2f{\widetilde f}_{s}\,{\widetilde k}_{V_0}k_{\perp}+{\widetilde f}_{s}^{\,\,2}k_{\perp}^2)-N_{\mu}^{2}k_{\perp}^2,
\end{equation}
\begin{equation}
E=\omega^2({\widetilde k}_{V_0}^{\,2}+k_{\perp}^2)-(f^2{\widetilde k}_{V_0}^{\,2}+2f{\widetilde f}_{s}\,{\widetilde k}_{V_0}k_{\perp}+{\widetilde f}_{s}^{\,\,2}k_{\perp}^2)-N_{T}^{2}k_{\perp}^2,
\end{equation}
\begin{equation}
F=2\omega^2({\widetilde k}_{V_0}^{\,2}+k_{\perp}^2)-N^2k_{\perp}^2.
\end{equation}
If $N_{\mu}^{2}$ and $\lambda_{\mu}$ vanish, we recover the result given in Eq. (\ref{Eq:TauNT}). In the general non-rotating case, we recover the result derived by \cite{Guoetal2023} where ${\widehat\tau}_{\Omega=0}=1/2\times k_{\perp}^{3}\times N^3/\omega^4\times\left(\kappa\times N_{T}^{2}/N^{2}+\lambda_{\mu}\times N_{\mu}^{2}/N^{2}+\nu\right)$, i.e. ${\rm Im}\left(k_V^2\right)=2\,{\widetilde k}_{V_0}\,\tau=k_{\perp}^{4}\times N^4/\omega^5\times\left(\kappa\times N_{T}^{2}/N^{2}+\lambda_{\mu}\times N_{\mu}^{2}/N^{2}+\nu\right)$. In the non-rotating case where $\left\{\lambda_{\mu},\nu\right\}\!<\!\!<\!\kappa$, we recover the result derived by \cite{Zahnetal1997} where ${\widehat\tau}_{\Omega=0}=1/2\times\kappa k_{\perp}^{3}\times N\,N_{T}^2/\omega^4$.
Finally, the global linear spatial damping rate is obtained:
\begin{equation}
\tau_{\rm G}\left(z\right)=\varepsilon\int_{z_0}^{z}\vert\,{\widehat\tau_{j}}\,\vert{\rm d}z'
\end{equation}
and we can treat the case of layers where both the entropy and chemical stratifications must be taken into account.

\subsection{Breaking of waves}

As we have seen in the main text, the breaking of GIWs is initiated by their convective and shear instabilities. Both instabilities are modified by the presence of chamical/$\mu$-gradients. First, depending on the sign of $\nabla_{\mu}$ (and thus of $N_{\mu}^{2}$) and the relative value of $N_{T}^{2}$, fingering convection ($\nabla_{\mu}<0$ and $N_{T}^{2}>0$), oscillatory double diffusive convection ($\nabla_{\mu}>0$ and $N_{T}^{2}<0$), or overturning convection are developing \citep[we refer the reader to][and references therein]{Garaud2021}. Specific work should be undertaken to study the potential breaking of GIWs in each of these configurations \citep{Worthemetal1983} and is beyond the current work. Next, the stability criterion for the vertical shear instability is modified \citep[e.g.][]{Maeder1995,Maeder1997,Maeder1996,TZ1997,PratL2014} and this must be taken into account when deriving the corresponding saturation vertical velocity ${\widehat W}_{\rm sat}^{\rm SWB}$ . This would be done once a robust prescription for the vertical shear instability with taking into account the full Coriolis acceleration and the chemical stratification would be obtained.     

\end{appendix}

\end{document}